\documentclass[aps,floatfix,showpacs,twocolumn]{revtex4}
\usepackage{amsmath}
\usepackage{graphicx}
\usepackage{amssymb}
\usepackage{color} % load color package

\definecolor{p}{rgb}{1, 0, 0}
\definecolor{y}{rgb}{0.3, 0.8, 0.3}

\begin{document}

\title{
Three-dimensional strong localization of matter waves by
   scattering from atoms in a lattice with a confinement-induced
   resonance }

\author{Pietro Massignan}

\affiliation{Niels Bohr Institutet, Universitetsparken 5, DK-2100 
Copenhagen \O,
Denmark,}

\affiliation{Laboratoire Kastler Brossel, \'Ecole Normale Sup\'erieure, 24 rue Lhomond,
75005 Paris, France.}

\author{Yvan Castin}

\affiliation{Laboratoire Kastler Brossel, \'Ecole Normale Sup\'erieure, 24 rue Lhomond,
75005 Paris, France.}

\date{\today}

\begin{abstract}
The possibility of using ultracold atoms to observe strong localization
of matter waves is now the subject of a great interest, as undesirable
decoherence and interactions can be made negligible in these systems.
It was proposed that a static disordered potential can be realized by trapping
atoms of a given species in randomly chosen
sites of a deep 3D optical lattice with no multiple occupation.
We analyze in detail the prospects of this scheme for observing
localized states in 3D for a matter wave of a different atomic species 
that interacts with the trapped particles 
and that is  sufficiently far detuned from  the optical lattice
 to be insensitive to it. 
We demonstrate that at low energy
a large number of  3D  strongly localized states can be produced for the matter wave,
 if the effective scattering length describing 
the interaction of the matter wave with a trapped atom is of the order
of the mean distance between the trapped particles. 
Such high values of the effective scattering length can be obtained by using a Feshbach
resonance to adjust the free space  inter-species 
scattering length and by taking advantage of confinement-induced resonances induced
by the trapping of the scatterers in the lattice. 
\end{abstract}
\maketitle

\section{Introduction}

The recent advances in the manipulation of ultracold gases have made it possible
to employ these systems to accurately simulate several non-trivial problems
in condensed matter physics. As examples, we may mention the exploration
of the BCS to BEC crossover \cite{BEC_to_BCS_crossover} 
and the superfluid to Mott-insulator
transition \cite{Greiner02}. Disorder plays an important role in
the theory of solid state, affecting in a substantial way the transport
properties of various systems. Special attention has in the past been
dedicated to studies of 
light propagation in strongly-scattering powders and electron transport
in the presence of impurities \cite{MesoscopicsBook,Lee85,Kramer93}.

It looks therefore interesting to introduce a controlled disorder
in the experiments with ultracold atoms, in order to provide a closer
modeling of realistic systems of condensed matter physics.
It was indeed predicted that atomic gases stored in optical lattices would be good
candidates to experimentally observe the effect of a disordered
or quasi-periodic potential on an interacting Bose gas or on interacting
Fermi-Bose mixture \cite{Lewen1,Lewen2}. First experimental results along this
research line have been recently reported, i.e.\ the observation of a Bose glass
in a quasi-periodic potential \cite{Inguscio} and the study of spatial
coherence properties
of an interacting Bose gas trapped in a lattice in presence of a disordered 
ensemble of fermionic atoms \cite{Sengstock}.

In this paper, we consider a variant of this line of research, 
that is the possibility of looking for genuine localized states 
of a non-interacting matter wave exposed to static disorder in continuous space. 
Localized states are stationary states with a square
integrable wave function at an energy where the classical motion is
not bounded spatially.
In a paper that dates back to the early years of quantum mechanics,
von Neumann and Wigner \cite{vonNeumann29}
showed that the  Schr\"odinger
equation can admit square integrable eigenstates embedded in the continuum
of states with energy higher than the maximum of the potential.
After the work of Anderson \cite{Anderson58},
it is expected that disordered potentials can generically lead
in 3D to a quantum phase transition, a macroscopic number of localized states
being present at low energy. Such a phase transition in 3D is not straightforward
to observe, as it is sensitive to decoherence and wave absorption
effects, and requires a mean free path of the wave $\mathit{l}$ smaller than its
wave length $\lambda$, as stated by the Ioffe-Regel criterion
\begin{equation}
\mathit{\lambda} \gtrsim l.
\label{eq:IRcrit}
\end{equation}

The study of localization of light is a well developed experimental subject:
strong localization of light has been reported in
semi-conductor powders \cite{Lagendijk}, 
and weak localization effects of light in a gas of cold atoms 
are the subject of an intensive experimental study \cite{Nice}.
On the contrary, for matter waves, no direct evidence of localization was
obtained in 3D. Matter waves made of ultracold atoms are good candidates
in this respect, due to their weak coupling to the environment and
to the possibility of tuning their interactions 
with a Feshbach resonance \cite{Feshbach}. An open problem is however to know if strong
enough disorder can be introduced in these gases to lead to 
reasonably short localization lengths in 3D.

A natural way to produce a disordered potential in atomic gases is
to use the speckle pattern of a laser beam \cite{Salomon_speckle}. Many experiments
on Bose-Einstein condensates in 1D random optical potentials
have very recently been reported \cite{Fallani05,Aspect05,Fort05,Schulte05},
and they provide evidence for disorder-related effects such as fragmentation
of the condensate, suppression of diffusion, frequency shifts and
damping of collective oscillations. Theoretically, these effects 
were discussed in \cite{Modugno05}.
Genuine strong localization
in 1D, in the non-interacting regime, has not been reported yet in
these experiments, and the implementation of the disordered optical
potential in 3D remains to be done. Also the theoretical analysis
of matter wave localization in a speckle pattern, recently performed in 2D
\cite{Delande}, has not been done in a detailed way in the 3D strong
localization regime.

\begin{figure}
\includegraphics[bb=10bp 100bp 360bp 285bp, clip, width=\columnwidth]{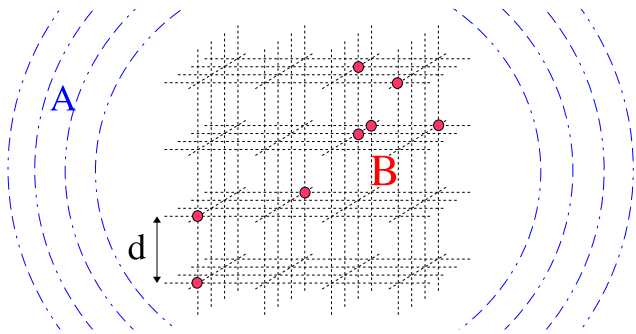}

\caption{\label{cap:Sketch}(Color online)
Sketch of the model of disorder considered here: 
a matter wave (A, blue)
scatters on randomly distributed B particles (red), each occupying
the vibrational ground state of a node in a 3D optical lattice (here
the average occupancy $p$ is 0.1). The lattice does not act on
the matter wave A.}
\end{figure}

An alternative method to realize a disordered potential was proposed
in \cite{Gavish05}: a matter wave, made of atoms of a species $A$,
scatters off a set of atoms of another species $B$ that are
trapped in randomly chosen sites of a deep optical lattice 
with  no multiply occupied sites 
(see Fig.~\ref{cap:Sketch}). 
As we will choose the lattice
to be very far detuned for the species $A$, the matter wave moves
unaffected through the optical lattice, and interacts only with the
$B$ atoms: this excludes classical localization effects in local
potential minima.
The disorder can be made very strong, since (i)
the correlation length of the disorder can be as small as $0.5\mu$m
(the spatial period of the optical lattice), and (ii) the scattering cross
section of the matter wave off a single $B$ atom can be made as high as
allowed by quantum mechanics (the so-called unitary limit) by use
of a $A-B$ interspecies Feshbach resonance,
making it possible
to dramatically reduce the mean free path of the matter wave. Furthermore,
as we shall take advantage of, this model allows a straightforward
exact numerical analysis even in 3D, when $B$ atoms are modeled as
fixed point-like scatterers, as is known for light waves \cite{Rusek95,Mandonnet}. 

It is the 3D version of this scheme that we analyze in this paper.
After the detailed presentation of our model and its practical implementation given in Sec.~\ref{sec:Our-proposal},
we show numerically in Sec.~\ref{sec:Localized-states} that it leads to the appearance of a large number
of localized states for a range of parameters accessible in present
experiments, provided that the effective coupling between the matter
wave and a single scatterer is tuned to a value of the order
of the mean scatterer separation. Section \ref{sec:Effective} is
dedicated to the quantitative description of the scattering between
the matter wave and a single trapped scatterer, and to the discussion
of the confinement-induced resonances thereby arising: we show
indeed that large enough effective coupling constants can be
obtained. Our conclusions are presented in Sec.~\ref{sec:Conclusions},
including a discussion of a possible strategy of observation of strong
localization.

\section{Our model\label{sec:Our-proposal}}

\subsection{The disordered potential}

The scatterers are a set of identical particles, whose chemical species
and quantum numbers will be indexed by the letter $B$, randomly occupying
(with filling factor $p<1$) the sites of a 3D cubic optical lattice.
The potential seen by the $B$ atoms is produced by a superposition
of three laser standing waves of common intensity and mutually orthogonal
linear polarizations along the $x$, $y$ and $z$
axes,\begin{equation}
V^{B}(\mathbf{r})=V_{0}^{B}[\sin^{2}(k_{L}x)+\sin^{2}(k_{L}y)+\sin^{2}(k_{L}z)],\label{eq:OpticalLattice}\end{equation}
 where $V_{0}^{B}>0$ is the modulation depth of the lattice and $k_{L}=2\pi/\lambda_{L}$
is the laser wavenumber. We shall denote the lattice spatial period
along each axis by $d=\lambda_{L}/2=\pi/k_{L}$. Multiple occupation
of a lattice well is assumed to be absent, by use of polarized fermions,
or by creation of vacancies in a unit occupancy Mott phase state \cite{Greiner02},
or simply by choosing $p$ sufficiently small to make it statistically
irrelevant. 

We choose the lattice depth $V_{0}^{B}$ to be much larger than the
recoil energy $E_{r}^{B}=\hbar^{2}k_{L}^{2}/2m_{B}$ of the $B$ atoms
so that the tunneling time of $B$ atoms from one lattice site to
another ($t_{\textrm{tunnel}}\approx 1.6\times10^{4}\hbar/E_{r}^{B}$
for $V_{0}^{B}=50E_{r}^{B}$) is negligible over the duration of the
experiment and the disordered spatial pattern of $B$ atoms is static \cite{tunnel}.

We also require that each $B$ atom is prepared in the vibrational ground
state of the local lattice micro-trap, which can be achieved in practice
by raising adiabatically the optical lattice on top
of a condensate cloud of atoms $B$ \cite{Greiner02}, or
by applying Raman laser cooling sideband techniques \cite{Hamann98,Perrin98} to
an optical molasses.
This condition is crucial to ensure that each $A-B$ scattering event is elastic when the
$A$ atoms have sufficiently low energy: indeed, energy conservation
guarantees that the $B$ atom is left in the vibrational ground state
after scattering with a $A$ atom of momentum $k$ if\begin{equation}
\frac{\hbar^{2}k^{2}}{2m_{A}}\ll\hbar\omega=2(V_{0}^{B}\, E_{r}^{B})^{1/2}\label{eq:ElasticScattering}\end{equation}
 where $\omega$ is the oscillation frequency of a $B$ atom in a
micro-trap. 

A last point is to ensure that spontaneous emission processes are
negligible for the $B$ atoms. In order to achieve large values of
$V_{0}^{B}$ with negligible heating of the trapped scatterers, we
require the lattice to be blue-detuned with respect to the strongest
transition of the $B$ atoms (in blue-detuned lattices, particles
are trapped in the minima of intensity of the stationary light field).
Including the Lamb-Dicke type reduction factor coming from the trapping
of $B$ atoms close to the nodes of the laser field, one gets the
fluorescence rate \begin{equation}
\Gamma_{\textrm{fluo}}^{B}=\Gamma_{B}\frac{V_{0}^{B}}{\omega_{L}-\omega_{B}}\frac{3k_{L}^{2}}{2m_{B}\omega}\end{equation}
 where $\omega_{L}-\omega_{B}$ is the atom-laser detuning and $\Gamma_{B}$
is the spontaneous emission rate of $B$ atoms. 

For the bosonic $^{87}$Rb isotope of rubidium ($\lambda_{B,D2}=780$nm,
$\lambda_{B,D1}=794.8$nm) and an optical lattice tuned at $\lambda_{L}=779$nm,
only at $1$nm to the blue of the strongest rubidium transition at
780nm, at the required lattice intensity $V_{0}^{B}=50E_{r}^{B}$
the tunneling time is $t_{\textrm{tunnel}}\approx 0.7$s \cite{tunnel} and the fluorescence
rate is $\Gamma_{\textrm{fluo}}^{B}\sim10^{-4}E_{r}^{B}/\hbar\sim3\mbox{s}^{-1}$,
allowing experimental times up to $300$ms.
The same calculation, taking for $B$ the fermionic isotope $^{40}$K of potassium
($\lambda_{B,D2}=766.5$nm, $\lambda_{B,D1}=769.9$nm) and an
optical wavelength $\lambda=765.5$nm leads to
$t_{\textrm{tunnel}}\approx 0.32$s and $\Gamma_{\textrm{fluo}}^{B}\sim 5\mbox{s}^{-1}$.

\subsection{Model Hamiltonian for the matter wave and its limitations}
\label{subsec:model_ham}

The matter wave to be strongly-localized is made of atoms of another
species, that we will label by $A$. We shall ignore interaction effects
among these $A$ atoms. One way to fulfill this condition in a real
experiment would be to take spin polarized fermionic atoms: $s$-wave
interactions are prohibited by the exclusion principle and $p$-wave
interactions are very weak at low energies in the absence of a $p$-wave
resonance.

The $A$ atoms experience interactions with the trapped species $B$.
At low incoming kinetic energy of a $A$ particle,
we model 
these interactions
by static contact potentials,
corresponding to infinitely-massive point-like scatterers, each located
at the center of a micro-well occupied by a $B$ atom: 
\begin{equation}
\mathcal{V}=\frac{2\pi\hbar^{2}a_{\mathrm{eff}}}{m_{A}}
\sum_{j=1}^{N}\delta(\mathbf{r}_{A}-\mathbf{r}_j)
\partial_{|\mathbf{r}_{A}-\mathbf{r}_j|}\left(|\mathbf{r}_{A}-\mathbf{r}_j|\ldots\right)\label{eq:Veff}\end{equation}
where the sum is taken over the $N$ scatterers.
The effective scattering length $a_{\rm eff}$ of a $A$ atom on a trapped $B$ atom,
when expressed in units of the harmonic-oscillator length 
$a_{\textrm{ho}}=\sqrt{\hbar/m_B\omega}$,
is a function of the
dimensionless ratios $m_B/m_A$ and $a/a_{\textrm{ho}}$, 
$a$ being the $A-B$ scattering length in free space.
The value of $a_{\mathrm{eff}}$ and the validity
condition of our model potential, Eq.~(\ref{eq:Veff}), will be given
in Sec.~\ref{sec:Effective}.

The $A$ atoms also experience the optical lattice potential, with
the same spatial dependence as in Eq.~(\ref{eq:OpticalLattice})
but with a different modulation amplitude $V_{0}^{A}$. We require
the optical lattice to be much closer to resonance with $B$ atoms
than with $A$ atoms, $\left|\omega_{L}-\omega_{A}\right|\gg\left|\omega_{L}-\omega_{B}\right|$,
such that $V_{0}^{A}$ will be much smaller than $V_{0}^{B}$. In
particular, we impose that \begin{equation}
|V_{0}^{A}|\ll E_{r}^{A}=\frac{\hbar^{2}k_{L}^{2}}{2m_{A}}\label{eq:shieldingForA}\end{equation}
 so that, in the absence of $B$ atoms, the $A$ atoms can be safely
considered as free. In this respect, a particularly promising combination
is given by fermionic $^{6}$Li for the species $A$ ($\lambda_{A}=671$nm)
and $^{87}$Rb for the species $B$: taking $\lambda_{L}=779$nm and
a laser intensity such that $V_{0}^{B}=50E_{r}^{B}$, one finds $V_{0}^{A}=-0.04E_{r}^{A}$.
If one takes for $B$ the fermionic $^{40}$K with $\lambda_L=765.5$nm,
one finds $V_{0}^{A}=-0.09E_{r}^{A}$ \cite{aPotassium}.

We shall therefore neglect the effect of the optical lattice on the
$A$ atoms and take as a model Hamiltonian for the matter wave: \begin{equation}
\mathcal{H}=\mathcal{H}_{0}+\mathcal{V}\,\,\,\,\,\,\,\,\mbox{with}
\,\,\,\,\,\,\,\,\mathcal{H}_{0}=-\frac{\hbar^{2}}{2m_{A}}\Delta_{\mathbf{r}_{A}}.\label{eq:model}\end{equation}
 We note in passing that in the original Anderson model the $A$ particles
were instead assumed to be in the tight-binding regime, so that the strong localization reported
in this paper is not {\sl stricto sensu} Anderson localization.

To end this section, we briefly discuss two effects not included in our model Hamiltonian
that may impose limitations in a real experiment. 
As we shall see, the production of localized states with a short localization length (of the order of the lattice
spacing $d$) requires the use of a large and positive value of $a_{\rm eff}\sim d$, obtained by a $A-B$
Feshbach resonance. As a consequence, the matter wave $A$ has a weakly bound state with a trapped atom $B$,
of spatial extension $\sim a_{\rm eff}$. A first undesired effect is therefore the formation of
such $A-B$ dimers. 

A first stage that may lead to a dimer production is during the Feshbach ramp of $a_{\rm eff}$ from
$\sim 0$ to $\sim d$. This may be avoided by using a ramping time longer than the inverse of the dimer binding frequency,
$2m_A a_{\rm eff}^2/\hbar \sim 30\mu$s for our previous example with lithium and rubidium.
Once $a_{\rm eff}$ is set to $\sim d$, one may fear that three-body collisions $A+A+B$ lead
to the formation of a dimer. For our model Hamiltonian Eq.~(\ref{eq:model}) the $A$ particles are an ideal
gas and the dimer formation does not happen: the trapped
$B$ particle, being replaced by a fixed scatterer with no degree of freedom, cannot mediate a $A-A$ interaction.
In the opposite limit where the $B$ scatterer is supposed to move freely, the rate of dimer formation per $B$ atom
is $\gamma_{\rm dim}= C_{\rm dim} (k_Fd)^8 E_r^A/\hbar$, 
where we used Eqs.~(11,12) of \cite{Dima_loss}, 
taking a dimer binding energy 
$\epsilon=\hbar^2/2\mu a_{\rm eff}^2$ with $\mu$ the $A-B$ reduced mass, $a_{\rm eff}=d$,
and assuming that the $A$ atoms are degenerate fermions of Fermi momentum $k_F$ with $k_F d < 1$; the constant $C_{\rm dim}$ 
is $6\times 10^{-5}$ for the $A={}^6$Li, $B={}^{87}$Rb case, resulting for $k_F d<1/2$ in a dimer formation rate much smaller than
e.g.\ the $B$ fluorescence rate $\Gamma_{\rm fluo}^B$ due to the lattice \cite{does_this_work}.
The calculation of the actual dimer formation rate in our model, taking into consideration the trapping
of the $B$ atoms, requires the solution of a 
three-body problem with no center-of-mass separability, which is beyond the scope of the present work.

A second undesired effect is the one of gravity.
If the lattice is arranged to be stationary in a free-falling
frame, this frame having initially an upward velocity component $V$ in the lab frame
and finally a downward velocity component $V$, the overall vertical motion of the lattice in such
a fountain-like experiment is less than 3 cm
for a total time of 150 ms. Longer times may be obtained if one compensates gravity, e.g.\, by a using
the inflexion point of the optical potential produced by a far-detuned Gaussian laser beam, or
by using  electro-optical potentials \cite{Morice}.
A drastic solution is of course to perform the experiment in a micro-gravity environment \cite{microgravity}.

\section{Localized states\label{sec:Localized-states}}

The Ioffe-Regel criterion Eq.~(\ref{eq:IRcrit}) is considered as a necessary condition to 
achieve strong localization \cite{Kramer93}. It is simple to check
that the setup that we consider can satisfy this criterion for
experimentally reasonable parameters. As
we will see in Sec.~\ref{sec:Effective}, the effective scattering
length presents confinement induced resonances
that allow one to reach the
unitary regime for the interaction of the matter wave with a trapped $B$ atom,
with the maximal cross-section $\sigma=4\pi/k^{2}$.
In this case, the Ioffe-Regel criterion reads:
\begin{equation}
kd\lesssim\left(4\pi p\right)^{1/3},
\end{equation}
which for a filling factor $p=0.1$ yields $k\lesssim 1/d=k_L/\pi$, 
achievable with sub-recoil laser cooling techniques, or with a low density
degenerate Fermi gas with a Fermi wavevector $k_F < k_L/\pi$.

Since the Ioffe-Regel criterion is not proved to be sufficient,
we numerically investigate in this section the possibility for the
disordered model Hamiltonian Eq.~(\ref{eq:model}) to lead to matter
wave localization. In particular, we shall find that the
unitary regime $a_{\rm eff}=\infty$ is not the most favorable one.

\subsection{How to find localized wavefunctions ? \label{sub:Direct-imaging-of}}

A criterion of strong localization
presented by Kramer and MacKinnon \cite{Kramer93} for electrons in a solid
consists in showing that, at the Fermi energy $E=E_F$,
off-diagonal elements of the resolvent $\mathcal{G}=(E+i0^{+}-\mathcal{H})^{-1}$
in real space decrease exponentially with the distance between the
two considered points. 

For our model Hamiltonian, the calculation of the matrix elements of the resolvent
is straightforward to implement numerically, using a technique well known
for scalar light waves in a gas of scatterers \cite{Rusek95,Mandonnet}.
These matrix elements are indeed given in presence of $N$ point-like scatterers by
\begin{eqnarray}
\left\langle \mathbf{r}\left|\mathcal{G}\right|\mathbf{r}^{\prime}\right\rangle &=&
g_{0}\left(\mathbf{r}-\mathbf{r}^{\prime}\right) +
\frac{2\pi\hbar^{2}}{m_{A}} \times\nonumber \\
&\times& \sum_{j,l}g_{0}\left(\mathbf{r}-\mathbf{r}_{j}\right)\left[M^{-1}\right]_{jl}g_{0}\left(\mathbf{r}_{l}-\mathbf{r}^{\prime}\right).\label{eq:r-G-rPrime}
\end{eqnarray}
Here $g_0$ is the propagator in free space of a particle of positive energy $E\equiv \hbar^2 k^2/2 m_A, k>0$:
\begin{equation}
g_{0}\left(\mathbf{r}-\mathbf{r}^{\prime}\right)\equiv\left\langle \mathbf{r}\left|\mathcal{G}_{0}\right|\mathbf{r}^{\prime}\right\rangle =-\frac{m_{A}}{2\pi\hbar^{2}}\frac{e^{ik\left|\mathbf{r}-\mathbf{r}^{\prime}\right|}}{\left|\mathbf{r}-\mathbf{r}^{\prime}\right|},\label{eq:G0Free}
\end{equation}
and we have introduced the $N\times N$ matrix $M$:
\begin{equation}
M=\frac{I}{a_{\mathrm{eff}}}+M^{\infty},
\label{eq:M}
\end{equation}
where $I$ is the identity matrix and $M^{\infty}$
a complex symmetric (not hermitian) matrix with elements defined by \begin{equation}
M_{jl}^{\infty}=\left\{ \begin{array}{cc}
\exp\left(ik\left|\mathbf{r}_{j}-\mathbf{r}_{l}\right|\right)/\left|\mathbf{r}_{j}-\mathbf{r}_{l}\right| & \textrm{ if }j\neq l,\\
ik & \textrm{ if }j=l.\end{array}\right.\end{equation}
The \emph{exact} calculation of the resolvent in coordinates space
is in this way reduced to the inversion of the 
$N\times N$ matrix $M$ \cite{link_with_T}.

We have implemented the criterion
by Kramer and MacKinnon for a variable energy $E$, and indeed we found an exponential decay
of $\left|\left\langle \mathbf{r}\left|\mathcal{G}\right|\mathbf{r}^{\prime}\right\rangle \right|^{2}$
for sufficiently low energies.
However, as shown in the appendix \ref{appen:kinnon}, this rapid decay is not a proof
of localization but may be due to the fact that $E$ is in a spectral gap of the system \cite{Ziman}. 

The most direct way to prove localization is to exhibit stationary
states that are `localized' inside the disordered potential, that
is with a wave function strongly peaked inside the scattering medium,
decreasing exponentially towards the borders of the scattering medium.
To this end, we use the fact that the wave function 
\begin{equation}
\phi(\mathbf{r};\mathbf{r}_{0})\equiv\mathrm{Im}\,\langle\mathbf{r}|\mathcal{G}(E+i0^{+})|\mathbf{r}_{0}\rangle,
\label{eq:the_def}
\end{equation}
when not identically zero, is an exact eigenstate of $\mathcal{H}$ with energy $E$, whatever the arbitrary location
of its center $\mathbf{r}_{0}$ \cite{qed}. Here we take $E>0$ so that $\phi(\mathbf{r};\mathbf{r}_{0})$
belongs to the continuum of the energy spectrum of $\mathcal{H}$,
like the scattering states. A useful expression of $\phi$ is then
\begin{equation}
\phi(\mathbf{r};\mathbf{r}_0) = A\,\mbox{Im}\,
\left[\frac{e^{i k|\mathbf{r}-\mathbf{r}_0|}}{|\mathbf{r}-\mathbf{r}_0|}+
\sum_{j=1}^{N} d_j \frac{e^{i k|\mathbf{r}-\mathbf{r}_j|}}{|\mathbf{r}-\mathbf{r}_j|}\right]
\label{eq:phi_jolie}
\end{equation}
where $A=-m_A/(2\pi\hbar^2)$ is a constant factor and
$d_j=-\sum_l (M^{-1})_{jl} \exp{(ik|\mathbf{r}_l-\mathbf{r}_0|)}/{|\mathbf{r}_l-\mathbf{r}_0|}$.
In practice, we choose $\mathbf{r}_0$ inside the scattering medium; to
see if $\phi$ is localized or not, one just has to compare the values of
$\phi$ inside and outside the scattering medium; one can even watch how
the modulus of $\phi$ decays inside the medium.

\subsection{Application of the proposed technique in 1D: analytical results}

We test the proposed technique in 1D, where analytical results can be obtained.
We assume that a quantum particle of mass $m$ interacts in 1D with
$N$ point-like scatterers that are randomly distributed
in the interval $x\in [-L/2,L/2]$.
Using the transfer matrix formalism, as detailed in the appendix \ref{appen:1D},
we can obtain exact expressions for the matrix elements of the resolvent $\psi(x)\equiv
\langle x|{\mathcal G}(E+i0^+)|0\rangle$ as a function of $x$. Assuming that there is no scatterer in
$x=0$, we obtain
\begin{equation}
\psi(0) = \frac{m}{i\hbar^2 k} \frac{(1+r_-)(1+r_+)}{1-r_- r_+}
\label{eq:psi1d_in}
\end{equation}
where $k>0$ is such that $E=\hbar^2 k^2/2m$, 
$r_-$ is the complex reflection coefficient for the set of scatterers in the half-space
$x<0$ oriented from $x=0$ to $x=-\infty$, and $r_+$ is the complex
reflection coefficient for the set of scatterers in the half-space $x>0$
oriented from $x=0$ to $x=+\infty$.
In the half-space $x>L/2$ we also have a simple expression:
\begin{equation}
\psi(x) = \frac{m}{i\hbar^2 k} \frac{t_+(1+r_-)}{1-r_- r_+} e^{ikx},
\label{eq:psi1d_out}
\end{equation}
where $t_+$ is the transmission coefficient of the set of scatterers in the half-space $x>0$
oriented from $0$ to $+\infty$. A similar expression holds for $x<-L/2$, see
appendix \ref{appen:1D}.

\begin{figure}[htb]
\begin{center}\includegraphics[clip,width=\columnwidth]{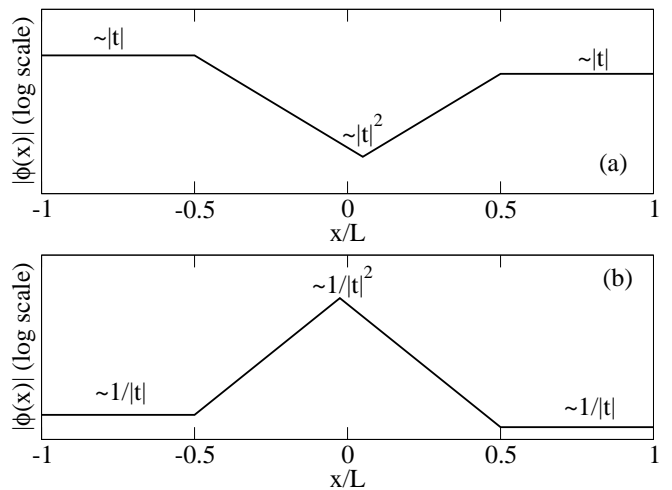}\end{center}

\caption{Schematic view of the behavior of the wavefunction $\phi(x)=\mbox{Im}\langle x|\mathcal{G}|0\rangle$
in presence of a 1D scattering medium of length $L$ much larger than the localization length $\xi$, so that
the modulus of the transmission coefficient for each half of the medium $|t_+|,|t_-|\sim |t|
\propto e^{-L/2\xi} \ll 1$.
(a) For a generic positive energy: the wavefunction decreases exponentially inside the medium, being
of modulus $\sim |t|$ out of the medium and $\sim |t|^2$ in the center $x=0$ of the medium.
(b) For specific values of the energy, the wavefunction is `localized' inside the medium: its modulus
decreases from $\sim 1/|t|^2$ in $x=0$ to $\sim 1/|t|$ for $|x|> L/2$.
\label{cap:schema_1d}}
\end{figure}

Using Furstenberg theorem \cite{furst}, we know that in the thermodynamic limit 
$L\rightarrow +\infty$, with
a fixed density of scatterers, the modulus of
the transmission coefficients tends exponentially to zero, 
\begin{equation}
|t_+| \sim |t_-| \sim |t| \propto e^{-L/2\xi}
\end{equation}
where $\xi$ is the localization length, so that the modulus of $r_+$ and $r_-$ tends to one.
We then find that the wavefunction
$\phi(x)=\mbox{Im}\,\psi(x)$ outside the scattering medium is of the order of
$|t|$.  To calculate $\phi$ inside the medium, we first take, to zeroth order in $|t|$,
$|r_+|=|r_-|=1$:  Eq.~(\ref{eq:psi1d_in}) then leads to a purely real $\psi(0)$ (of order unity), that is
to a vanishing $\phi(0)$. Going to next order, $|r_\pm|\simeq 1-|t_\pm|^2/2$ leads to
$|\phi(0)|\sim |t|^2$. For a generic value of $k$, the wavefunction $\phi(x)$ has therefore
the behavior depicted in Fig.\ref{cap:schema_1d}a.

However, for specific values $k_0$ of $k$, the above reasoning is incorrect.
Assume that $k$ is such that $r_+ r_-$ is a real and positive number.
Then the denominator $1-r_+r_-$ of Eqs.(\ref{eq:psi1d_in},\ref{eq:psi1d_out}) is extremely small,
of order $|t|^2$. This leads to $|\phi(x)|$ decreasing exponentially from $x=0$
outwards, from a value $\sim 1/|t|^2$ to a value $\sim 1/|t|$ \cite{bad_case}, as depicted in
Fig.\ref{cap:schema_1d}b. In this case, $\phi(x)$ corresponds to a `localized' state
inside the medium.  

Strictly speaking, since we
consider disorder over a finite region, $\phi(x)$ is
not localized: the exponential decrease of the envelope
stops outside the scattering medium, so that $\phi(x)$ is not square integrable.
The state $\phi(x)$ rather corresponds to a resonance, whose lifetime is of the order
of $\hbar$ over the energy width of the resonance.
In practice, $\phi$ can be considered as
localized if the resonance lifetime is much longer than the duration of the experiment.
Assuming that the phase of $r_+ r_-$ varies linearly with $k$ close to $k_0$,
one finds that this `localized' state is present on a narrow interval in $k$ 
of width $\propto |t|^2$, so that its energy width, or equivalently its inverse lifetime,
scales as
\begin{equation}
\Gamma \propto \frac{|t_+|^2+|t_-|^2}{2} \sim |t|^2 \propto e^{-L/\xi}.
\label{eq:width_1d}
\end{equation}

This illustrates how the proposed technique, defined in Eq.~(\ref{eq:the_def}),
gives access to localized wavefunctions.

\subsection{Application of the proposed technique in 3D: numerical results}

We now present numerical results
obtained for a single realization of a random potential obtained by a
Monte Carlo generation of the positions of a finite number $N$ of scatterers at the nodes
of a cubic lattice with a given filling factor $p$.

\begin{figure}
\begin{center}\includegraphics[clip, width=\columnwidth]{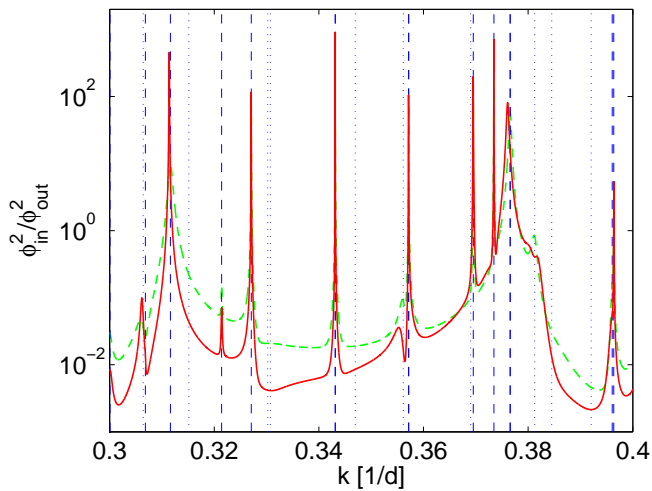}\end{center}

\caption{\label{cap:scan} (Color online)
Ratio of the values of $\phi^{2}$ in the center and outside the scattering medium 
(see text)
as a function of $k=(2m_{A}E)^{1/2}/\hbar$, with $a_{\mathrm{eff}}=d$,
for a given realization of the disorder with $N\approx900$ scatterers that occupy the nodes
of a cubic lattice with 21 sites per side, with a filling factor $p=0.1$.
The value of $\phi^2_\mathrm{out}$
 is calculated as explained in \cite{details} (red solid curve)
or with the extrapolation from the far field behavior, 
Eq.\ (\ref{ang_average}) with $r=R=20d$ (green dashed curve).
The energy intervals where the matter wave significantly penetrates
the scattering medium correspond to the narrow peaks in this figure.
We have checked that the wave function is actually spatially localized in
such an energy interval. The vertical lines mark the locations of the resonances obtained
by the spectral method: 
dashed lines for the very long lived resonances 
($\Gamma < 6\times 10^{-5} E_r^A/\hbar$) and dotted lines for the broader resonances. The 
temporal decay rate $\Gamma$ is obtained from Eq.~(\ref{eq:width}).
}
\end{figure}

In Fig.~\ref{cap:scan} we plot the ratio of the square of the amplitudes of $\phi$
inside and outside the scattering medium, 
for $N\simeq 900$ scatterers and a filling factor $p=0.1$.
In order to avoid a choice that might pick a node of $\phi(\mathbf{r})$,
we plot $\phi_{\textrm{in}}^{2}/\phi_{\textrm{out}}^{2}$,
where both numerator and denominator are averaged over a few points \cite{details}.
The graph
 reveals that the phenomenology is similar to the 1D case:
one has generically $\phi_{\rm in}^2\ll \phi_{\rm out}^2$, except for narrow energy intervals,
corresponding to the peaks in the figure,
where the matter wave can significantly penetrate the scattering medium. 
We have verified for a large number of peaks that the wavefunction $\phi$ is indeed
`localized' inside the medium. 

\begin{figure}
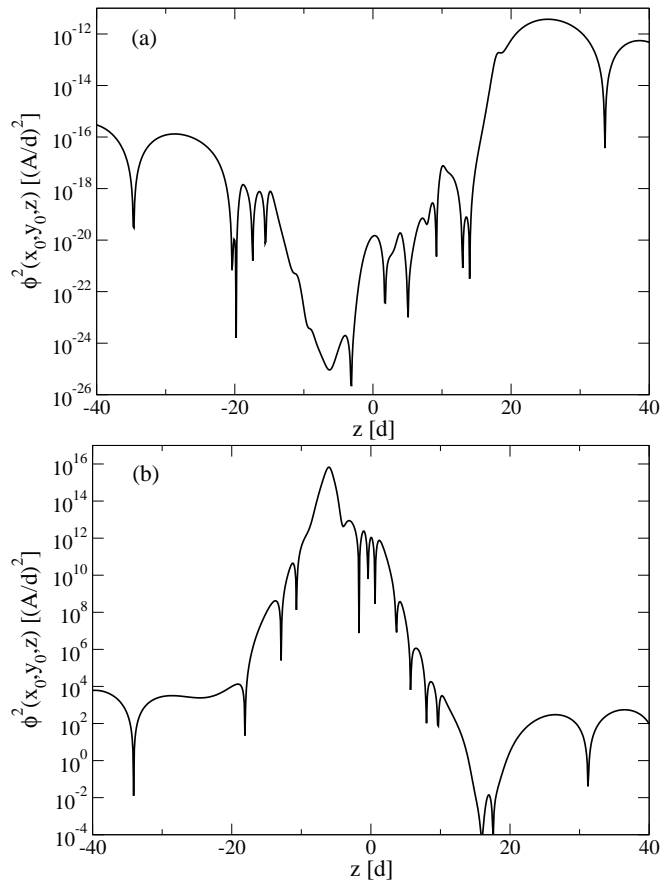

\begin{center}
\includegraphics[clip,width=\columnwidth]{./fig4a.eps} \\
\includegraphics[clip,width=\columnwidth]{./fig4b.eps}
\end{center}

\caption{\label{cap:etats3d}
Plot of $\phi(\mathbf{r})^2$ along
the straight line passing through $\mathbf{r}_0$
and parallel to $z$ axis, for two values of $k$, (a) a generic value $k=0.3/d$
and (b) a value $k=0.350134274724/d$ corresponding to a peak for $\phi_{\rm in}^2/\phi_{\rm out}^2$
as a function of $k$ (in the spirit of Fig.\ref{cap:scan}).
Note the similarity with the 1D case sketched in  Fig.\ref{cap:schema_1d}.
In (a) the position $\mathbf{r}_0$ is close to the center of the scattering
medium: $\mathbf{r}_0=(d/2,-d/2,-d/2)$; in (b) it is close to the `center' of
the localized state: $\mathbf{r}_0 =(15d/2,-3d/2,-11d/2)$.
The effective scattering length is given by 
$d/a_{\rm eff}=1.20530122302$.
To get a clear evidence of the exponential decay of $\phi$,
we used a larger scattering medium
than in Fig.~\ref{cap:scan}: $N \simeq 3400$ atoms on the lattice within a sphere of radius 
$20 d$, with an occupation probability $p=0.1$.
}
\end{figure}

We illustrate this phenomenology for a generic value of $k$, and for one that corresponds
to a peak in the function $\phi_{\rm in}^2/\phi_{\rm out}^2$
taking now a larger number of scatterers $N\simeq 3400$.  
As can be seen in Fig.~\ref{cap:etats3d}, 
in the generic case $\phi(\mathbf{r})$ decays essentially exponentially
when entering the medium, whereas for the specific
value of $k$ the wavefunction decays essentially exponentially
from the medium center towards the outside. What is the associated
localization length $\xi$~? If one takes $\phi\propto \exp(-|\mathbf{r}-\mathbf{r}_0|/\xi)$,
one obtains from a fit the estimate $\xi\sim d$ \cite{what_is_xi}.

\subsection{Comparison with a spectral technique}

\begin{figure}
\begin{center}\includegraphics[clip,width=0.9\columnwidth]{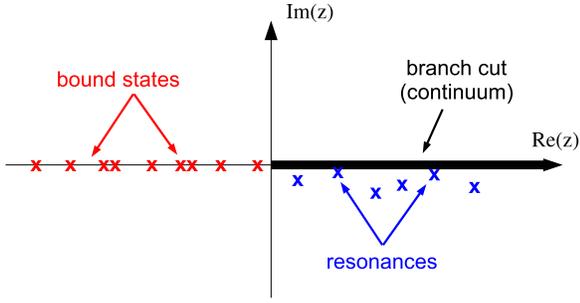}\end{center}

\caption{(Color online) Analytic properties of the resolvent $\mathcal{G}(z)=1/(z-\mathcal{H})$ in the complex plane
$z$. The resolvent has poles on the negative side of the real axis, corresponding to
bound states, and a branch cut on the positive side of the real axis, corresponding
to the continuum. The analytic continuation of the resolvent across the branch cut
from $\mbox{Im}\, z>0$
to $\mbox{Im}\, z<0$ may present a discrete set of poles in the fourth quadrant: the associated
states are the resonances of the system.
\label{cap:Poles-of-the-resolvent}}
\end{figure}

In this subsection, we adopt a different point of view on localization \cite{Imry}, based
on the properties of the analytic continuation of the resolvent $\mathcal{G}(z)=1/(z-\mathcal{H})$ from the half-plane
$\mathrm{Im}\, z >0$ to the half-plane $\mathrm{Im}\, z <0$, across the branch cut $\mbox{Im}(z)=0, 
\mbox{Re}\,(z)>0$ corresponding to the continuous spectrum of the Hamiltonian.
This analytic continuation can have poles in $z=E_0-i\hbar \Gamma/2$, where $E_0$ and $\Gamma$ are
positive (see Fig.~\ref{cap:Poles-of-the-resolvent}): these poles correspond to resonances. The lifetime
of the state associated to the resonance is given by $\Gamma^{-1}$.

The key property of localized states is that they correspond to resonances
with a decay rate $\Gamma$ that tends to zero in the limit of an infinite extension
of the disorder, so that they become in this limit square integrable stationary states
of the Hamiltonian. 
We expect that the poles associated to these narrow resonances will 
leave signatures on the real axis, in the form of 
eigenvalues of the matrix $M$ with vanishing real part and tiny imaginary part, for an energy $E=\hbar^{2}k^{2}/2m_{A}$
close to the real part $E_0$ of the poles. 
 As opposed to the poles of $\mathcal{G}$,
the eigenvalues of $M$ can be calculated in a straightforward manner,
and in Fig.~\ref{cap:scan} we show with vertical lines the values of $k$ for
which the matrix $M$ has a purely imaginary eigenvalue: the ones with the smallest
imaginary parts (dashed vertical lines) are in good correspondence with the narrow peaks in $\phi^2_{\rm in}
/\phi^2_{\rm out}$.

\begin{figure}
\begin{center}\includegraphics[width=\columnwidth,keepaspectratio]{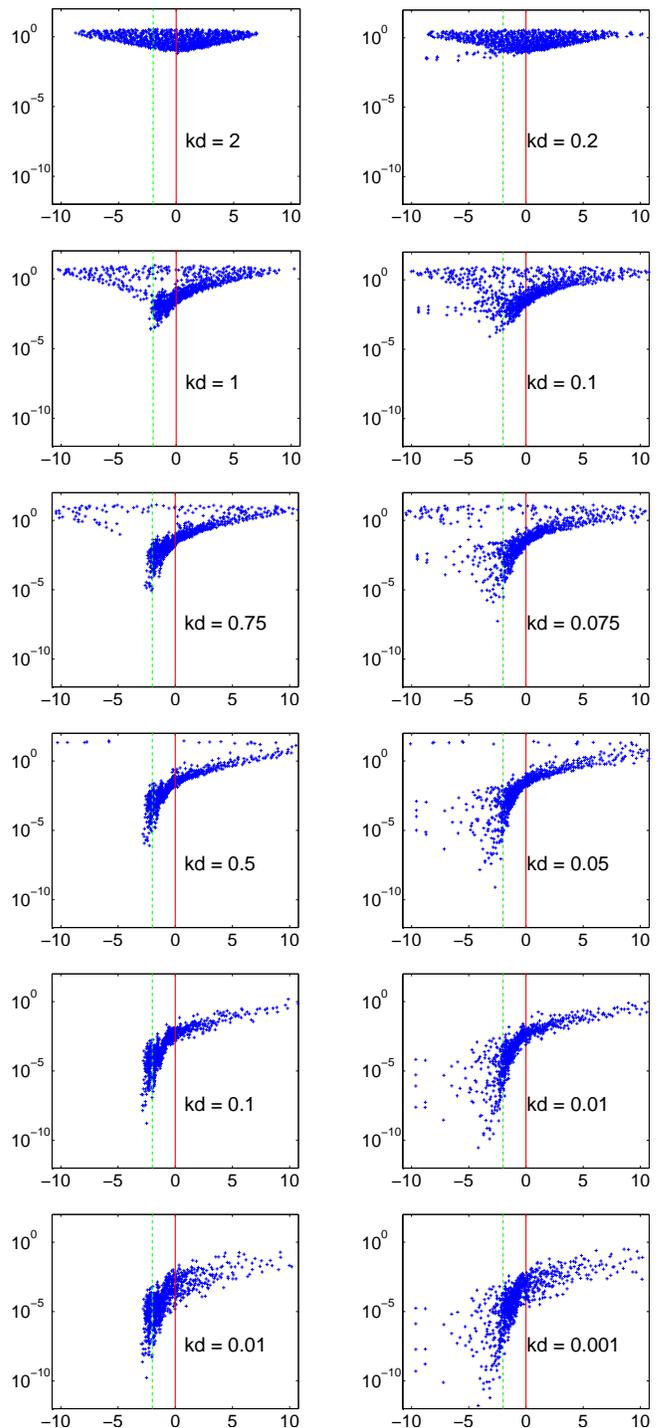}\end{center}

\caption{(Color online)
Representation in the complex plane of the
eigenvalues $m^{\infty}$ of $M^{\infty}$ for different values of the wave number $k$ of the matter wave.
The eigenvalues of $M$ are then simply deduced from these eigenvalues by a
shift of $1/a_{\rm eff}$ along the real axis.
Left column: cubic lattice with 21 sites/side, $p=0.1$ 
and 872 scatterers; the inverse  mean distance between scatterers is
$p^{1/3}/d \simeq 0.46/d$ (same realization of disorder as in 
Fig.~\ref{cap:scan}). Right column: cubic lattice with 209
sites/side, $p=10^{-4}$ and 878 scatterers; 
the inverse  mean distance between scatterers is $p^{1/3}/d \simeq 0.046/d$. 
The green dashed lines mark the values $\textrm{Re}(m^{\infty})=-2p^{1/3}/d$.
The real axis is in units of $p^{1/3}/d$, and common values of $kd/p^{1/3}$ are taken in both columns, 
so as to reveal a possible universality in the low
$p$ limit.  The imaginary axis is in units of $k$, as justified by Eq.~(\ref{eq:decay_to_m0}).
\label{cap:Eigenvalues-of-M}}
\end{figure}

Now, for a value of $k$ within the width of a peak in Fig.\ref{cap:scan}, 
we give an analytical argument relating the spatial decay of $\phi(\mathbf{r})$ to the presence
of a tiny eigenvalue $m_0$ of $M$ of modulus much smaller than the other eigenvalues.
In the large $r$ limit, Eq.~(\ref{eq:phi_jolie}) reduces to the far field expression 
\begin{equation}
\phi(\mathbf{r};\mathbf{r}_0) = A\, \mbox{Im}\, \left[\frac{e^{ikr}}{r} \left(
e^{-i k \mathbf{n}\cdot \mathbf{r}_0}+\sum_{j=1}^{N} d_j e^{-i k \mathbf{n}\cdot \mathbf{r}_j}\right)\right],
\label{eq:far_field}
\end{equation}
with $\mathbf{n}=\mathbf{r}/r$.
As shown in the appendix \ref{appen:magique}, 
the sum over $j$ in the right hand side is typically
$(\mbox{Im}\, 1/m_0^*)^{1/2}$ times larger
than the first term so that the angular average of $\phi^2$ is given,
apart from oscillating terms $\sim e^{2i k r}/r^2$, by
\begin{equation}
\langle \phi(\mathbf{r};\mathbf{r}_0)^2\rangle_{\mathbf{n}} \sim A^2\, \frac{\vec{d}\,^*\cdot (\mbox{Im}\, M)\vec{d}}{2 kr^2}.
\label{ang_average}
\end{equation}
This allows to estimate $\phi_{\rm out}$ by extrapolating this far field expression down to $r=R$,
where $R$ is a distance of order the size of the scattering medium \cite{stronger},
see the green dashed curve in Fig.\ref{cap:scan}.
We now need to estimate the wavefunction inside the medium. Close to the scattering center located in
$\mathbf{r}_j$, one finds $\phi\sim A\, \mbox{Im} \, d_j/|\mathbf{r}-\mathbf{r}_j|$. This suggests that the $d_j$ are also
localized, in the sense that they decrease rapidly from a center in the medium outwards: this we have checked
numerically. Then, averaging spatially over a small volume $l^3$, $l$ being smaller than the mean distance between
scatterers,
one finds that $\phi_{\rm in}^2$ is at most of the order of 
$(A/l)^2\mbox{max}_j |\mbox{Im}\ d_j|^2$, a value reached when $\mathbf{r}_0$ is close to the
`center' of the localized wavefunction. 
We reach the estimate
\begin{equation}
\frac{\phi^2_{\rm in}}{\phi^2_{\rm out}} 
\sim \frac{R^2}{l^2} \, \frac{\mbox{max}_j |\mbox{Im}\, d_j|^2}{\vec{d}\,^*\cdot (\mbox{Im}\, M/k) \vec{d}}\quad.
\label{eq:est_int}
\end{equation}
In the right hand side, the first factor has a geometrical origin whereas the second one is sensitive to
matter wave interference effects due to multiple scattering on the $B$ atoms: since the matrix elements of
$\mbox{Im}\, M/k$ are of the order of unity, only interference effects can indeed lead to a very small
expectation value of this matrix.
Using arguments detailed in the appendix \ref{appen:magique}, we ultimately arrive at
\begin{equation}
\frac{\phi^2_{\rm in}}{\phi^2_{\rm out}} \sim \frac{R^2}{l^2} \, \mbox{Im}\,\left[\frac{k}{m_0^*}\right],
\label{eq:decay_to_m0}
\end{equation}
which links the spatial decay of the wavefunction to the smallness of
an eigenvalue of the matrix $M$.
A useful application of this formula is to give the shape of the resonances in Fig.\ref{cap:scan}. 
We linearize the $k$ dependence of $m_0$ around the value $k_0$ such that $m_0$ is purely imaginary:
$m_0(k)\simeq  \beta(k-k_0) + i\alpha$, where $\alpha>0$. Anticipating Fig.~\ref{cap:Eigenvalues-of-M},
it appears that the imaginary part of an eigenvalue of $M^\infty$, when tiny for $k=k_0$,
remains tiny for even lower values of $k$, so that the derivative of $\mbox{Im}\,m_0$
is also tiny;  since the real part varies on the contrary over an interval of
width $1/d$, its derivative is not extremely small: $\beta$ is essentially real \cite{second_argument}.
Eq.~(\ref{eq:decay_to_m0}) then leads to a Lorentzian shape 
of the peaks in Fig.\ref{cap:scan}, in agreement with the numerics. Since the peaks
have a width much narrower than $k_0$, this also leads to a
Lorentzian dependence with the energy $E=\hbar^2 k^2/2m_A$, with a half-width at half maximum \cite{about_sign}:
\begin{equation}
\frac{\hbar\Gamma}{2} \simeq \frac{\hbar^2 k_0}{m_A} \, \frac{\mbox{Im}\, m_0(k_0)}{\mbox{Re}\, m_0'(k_0)}
\propto \mbox{Im}\, m_0.
\label{eq:width}
\end{equation}
When combined to Eq.~(\ref{eq:decay_to_m0}), this leads to a formula nicely relating
the inverse lifetime $\Gamma$ of the localized state to its spatial decay:
\begin{equation}
\Gamma \propto \frac{\phi_{\rm out}^2}{\phi_{\rm in}^2}(k=k_0).
\label{eq:width_3d}
\end{equation}
If one then assumes a state with a wavefunction localized close to the center of the medium
and decaying exponentially as $e^{-r/\xi}$ away from the center,  Eq.~(\ref{eq:decay_to_m0}) leads
to $\mbox{Im}\, m_0 \propto e^{-L/\xi}$, where $L$ is the diameter of the scattering medium, and
Eq.~(\ref{eq:width_3d}) leads to a resonance energy width $\Gamma\propto e^{-L/\xi}$,
thus obeying the same scaling as in 1D, see Eq.~(\ref{eq:width_1d}).

An important practical consequence of the present spectral approach is to give at once the range of values of $a_{\rm eff}$ 
for which one can hope to have localized states. For a given value of $k$,
$M$ will have a tiny purely imaginary eigenvalue if $1/a_{\rm eff}$ is opposite to
the real part of an eigenvalue of $M^{\infty}$ with a very small imaginary part, see Eq.~(\ref{eq:M}).
In Fig.~\ref{cap:Eigenvalues-of-M} the eigenvalues of $M^{\infty}$  are shown as points
in the complex plane, for various values
of the incoming wave number $k$ and for two realizations of
the disorder with widely different densities of scatterers. 
We observe
that, as the incoming wave number $k$ decreases below the inverse
of the mean distance between the scatterers $p^{1/3}/d$,      
many eigenvalues acquire an extremely small imaginary
part and accumulate in the region 
$\mbox{Re} \,m^\infty\sim -2p^{1/3}/d$
 (see the green dashed lines
in Fig.~\ref{cap:Eigenvalues-of-M}), similarly to 
earlier calculations for light waves  \cite{Rusek00}.

\begin{figure}[htb]
\begin{center}
\includegraphics[width=\columnwidth]{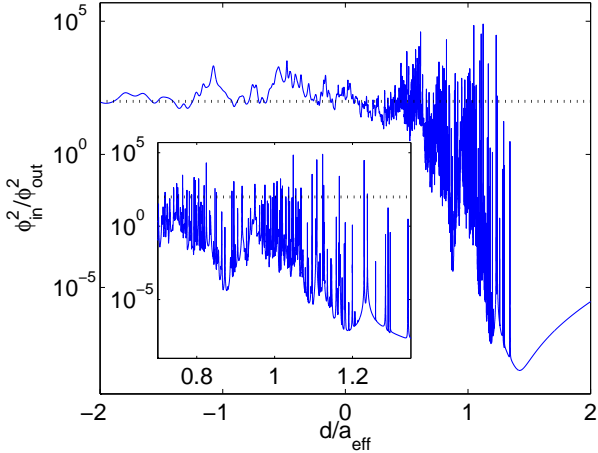}
\end{center}
\caption{(Color online)
For a fixed positive energy, corresponding to $k=0.35/d$,
ratio of the values of $\phi^{2}$ in the center and outside the scattering medium as
a function of the effective scattering length $a_{\rm eff}$ \cite{details}. 
The dashed line gives the result for $a_{\rm eff}=0$. The same realization of disorder is 
used as in Fig.\ref{cap:scan}. The inset is a magnification of the region $a_{\rm eff}\sim d$.
\label{cap:fonc_aeff}}
\end{figure}

In a matter wave experiment, this suggests to tune $a_{\mathrm{eff}}^{-1}$ to a value
close to $2p^{1/3}/d$ (as we have done in Fig.\ref{cap:scan}): this might be achieved in practice
by using a Feshbach resonance, as shown in the
next section. In this way, as $k$ decreases below $2p^{1/3}/d$, one obtains a very large sequence of
values of $k$ such that $M$ has a tiny purely imaginary eigenvalue, that
is one may have access to a large number of localized states. This is illustrated in Fig.\ref{cap:fonc_aeff}.

Finally, the representation in Fig.\ref{cap:Eigenvalues-of-M} is also useful to
understand what happens in the 
low energy limit,  $k\rightarrow 0$
It shows that, for each eigenvalue $m^\infty$ of the 
matrix $M$, $\mbox{Re}\,m^\infty$ and $\mbox{Im}\,m^\infty/k$ have a finite limit 
\cite{perturb}; a numerical inspection reveals that some of the eigenstates of $M^\infty(k=0)$, having
a tiny value of $\lim_{k\rightarrow 0}\mbox{Im}\,m^\infty/k$, are localized.
Can these localized states in the zero energy limit be accessed in a real experiment~? For a given realization
of disorder, this would require that $a_{\rm eff}$ be tuned exactly to one of the corresponding values of
$-1/\mbox{Re}\,m^\infty$, which is unrealistic.

In the same way, it would be very difficult
to adjust $a_{\rm eff}$ to hit, at a {\it given} value of $k$, one of the peaks in Fig.\ref{cap:fonc_aeff}.
Fortunately, in a real experiment, $a_{\rm eff}$ is fixed and a broad interval of $k$ 
can be populated by the atomic wavepackets;
then, Fig.\ref{cap:Eigenvalues-of-M}
shows that the real parts of the eigenvalues of $M^\infty$ with tiny imaginary parts are increasing
functions of $k$, so that they have a chance to cross the value $-1/a_{\rm eff}$ for some value of $k$ within
the experimentally populated interval, and to lead 
to a peak in Fig.\ref{cap:scan} and to an observable localized state in the experiment.

\begin{figure}[htb]
\begin{center}
\includegraphics[width=\columnwidth]{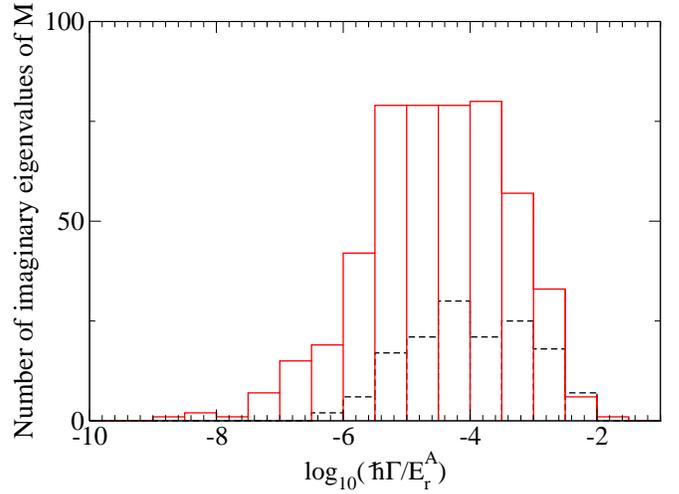}
\end{center}
\caption{(Color online)
For a single realization of disorder,
histogram giving the number of imaginary eigenvalues of $M$
per class of inverse lifetimes $\Gamma^{-1}$.
The filling factor of the lattice is $p=0.1$ and the effective scattering length
is $a_{\rm eff}=d$. Black dashed histogram: $N=872$ scatterers within a cube
with 21 sites per side. Red solid line histogram: $N=2985$ scatterers within a cube
with 31 sites per side. These histograms were constructed by a dichotomy search
of the values $k_0$ of the momentum $k \in [0,1/d]$ such that the matrix $M$
has a purely imaginary eigenvalue; the associated lifetime was then calculated
with Eq.~(\ref{eq:width}). Eigenvalues with negative values of $\Gamma$ are of course not included (see \cite{about_sign}). 
The decay rate $\Gamma$ is given in units of the recoil angular frequency
$E_r^A/\hbar=\hbar k_L^2/2m_A$ of the species $A$. In practice, only the eigenvalues with small enough
values of $\Gamma$ are expected to produce an observable resonance 
in $\phi_{\rm in}^2/\phi_{\rm out}^2$ as
a function of $k$, see Fig.\ref{cap:scan}.
If the matter wave is made of atoms of $^{6}$Li, a duration of the experiment
of 0.3s corresponds to a minimal observable inverse lifetime of $\Gamma\sim 10^{-5} E_r^A/\hbar$
for an optical lattice laser wavelength of 779 nm.
\label{cap:histo}}
\end{figure}

To conclude this section, it is useful to discuss the number of localized states
that can be supported by a finite size scattering medium, to have an idea of the number of
atoms that may populate these localized states in a real experiment, and to have an
estimate of the number of states that can be considered as localized for a given
duration of the experiment.
To this end, we present in Fig.\ref{cap:histo} an histogram giving the number
of purely imaginary eigenvalues of $M$ per class of inverse lifetimes $\Gamma$.
It is apparent on this figure how an increase of the volume of the scattering medium
(here by a factor $\sim 3.3$) leads to both an increase in the total
number of localized states (for a given lifetime) and to the appearance of a tail
of states of significantly longer lifetimes (here by about two orders of magnitude).

\section{Effective interaction\label{sec:Effective}}

We will solve here the two-body problem of a free atom $A$ scattering
on a single trapped particle $B$ prepared in the ground state of a harmonic oscillator
potential.
In the limit of vanishing incoming kinetic energy of the $A$ particle,
this gives access to the effective scattering length introduced
in Eq.~(\ref{eq:Veff}), which is the key parameter of our model
Hamiltonian with disorder.

\subsection{Calculation of the $A$-$B$ scattering amplitude}

The Hamiltonian for a free $A$ particle and a harmonically trapped $B$ one is,
in the absence of $A$-$B$ interaction,
\begin{equation}
H_{0}=-\frac{\hbar^{2}\Delta_{A}}{2m_{A}}-\frac{\hbar^{2}\Delta_{B}}{2m_{B}}+\frac{1}{2}m_{B}\omega^{2}r_{B}^{2}.
\label{eq:twobodyH_0}\end{equation}
It admits the $s$-wave solution  \cite{why_s}
\begin{equation}
\psi_{0}(\mathbf{r}_{A},\mathbf{r}_{B})=\frac{\sin(kr_{A})}{kr_{A}}\phi_{0}(r_{B}),
\label{eq:swavePsi_o}
\end{equation}
which represents a spherical matter wave ($A$) of wave vector
$k$ and a particle ($B$) in the ground state of the harmonic potential,
$\phi_{0}(r_{B})=\exp(-r_{B}^{2}/2a_{\mathrm{ho}}^{2})/\left(\sqrt{\pi}a_{\mathrm{ho}}\right)^{3/2}$,
with the harmonic oscillator length $a_{\rm ho}=(\hbar/m_B\omega)^{1/2}$.
The state $\psi_0$ has an energy 
\begin{equation}
E=\frac{\hbar^{2}k^{2}}{2m_{A}}+\frac{3}{2}\hbar\omega.
\end{equation}
We model
the $A$-$B$ interaction with a regularized contact potential, which leads
to the full Hamiltonian:
\begin{equation}
H=H_{0}+g\delta(\mathbf{r}_{A}-\mathbf{r}_{B})\frac{\partial}{\partial|\mathbf{r}_{A}-\mathbf{r}_{B}|}\left(|\mathbf{r}_{A}-\mathbf{r}_{B}|\ldots\right).\label{eq:twobodyH}\end{equation}
Here, the derivative is taken for a fixed value of the center of mass
position $\mathbf{R}$ of the two particles,
the coupling constant $g$ is expressed in terms of the reduced
mass $\mu=m_{A}m_{B}/(m_{A}+m_{B})$ and the free-space scattering
length $a$ (relative to the $A$-$B$ interaction in the absence of the trapping
potential) by \begin{equation}
g=\frac{2\pi\hbar^{2}a}{\mu}.\label{eq:g}\end{equation}
The Schr\"odinger equation $H\psi=E\psi$ can be reformulated equivalently
in the integral form\begin{equation}
\psi(\mathbf{r}_{A},\mathbf{r}_{B})=\psi_{0}(\mathbf{r}_{A},\mathbf{r}_{B})+g\int\textrm{d}\boldsymbol{\rho}\, G_{E}(\mathbf{r}_{A},\mathbf{r}_{B};\boldsymbol{\rho},\boldsymbol{\rho})\psi_{\mathrm{reg}}(\rho)\label{eq:ScatteringSolution}\end{equation}
in terms of the two-particle retarded Green's function for the non-interacting
Hamiltonian\begin{equation}
G_{E}=\frac{1}{E+i0^{+}-H_{0}}\end{equation}
and the regularized part of the two-particle wave function,\begin{equation}
\psi_{\mathrm{reg}}(\mathbf{R})=\left.\frac{\partial}{\partial|\mathbf{r}_{A}-\mathbf{r}_{B}|}\left[|\mathbf{r}_{A}-\mathbf{r}_{B}|\psi\left(\mathbf{r}_{A},\mathbf{r}_{B}\right)\right]\right|_{\mathbf{r}_{A}=\mathbf{r}_{B}}\end{equation}
where $\mathbf{R}=(m_A\mathbf{r}_{A}+m_B\mathbf{r}_{B})/(m_A+m_B)$ 
and $\psi_{\mathrm{reg}}(\mathbf{R})=\psi_{\mathrm{reg}}(R)$
since we consider $s$-wave scattering only. Inserting Eq.~(\ref{eq:ScatteringSolution}) into the definition of
$\psi_{\mathrm{reg}}$,
we find that the regularized part of the wave function satisfies an
equation of the form\begin{equation}
\tilde{\psi}_{\mathrm{reg}}=\frac{I}{I-g\hat{O}}\tilde{\psi}_{0},\label{eq:IntegralEquation}\end{equation}
where $\tilde{\psi}(R)=R\psi(R)$, $\tilde{\psi}_0(R)=R\phi_0(R) \sin(kR)/(kR)$,
and $\hat{O}$ is an integral operator independent of the scattering length $a$.
The detailed derivation of this equation and the explicit form of
$\hat{O}$ are presented in Appendix \ref{appen:Inte}.

As we now show, the knowledge of $\tilde{\psi}_{\rm reg}$ directly leads to the value of the scattering
amplitude $f_k$ for the $A$ wave.
Expanding $G_{E}$ on the basis of eigenstates $|\mathbf{k}_A,\mathbf{n}\rangle$ of $H_{0}$,
and projecting in position space, one is able to calculate the integral over the wavevector
$\mathbf{k}_A$ so that one is left with a sum over the vibrational states of the $B$ particle:
\begin{multline}
G_{E}(\mathbf{r}_{A},\mathbf{r}_{B};\boldsymbol{\rho},\boldsymbol{\rho})=\frac{2m_{A}}{\hbar^{2}}\left[-\frac{e^{ik\left|\mathbf{r}_{A}-\boldsymbol{\rho}\right|}}{4\pi\left|\mathbf{r}_{A}-\boldsymbol{\rho}\right|}\phi_{0}(\mathbf{r}_{B})\phi_{0}(\boldsymbol{\rho})\right.\\
\left.-\sum_{\mathbf{n}\neq0}\frac{e^{-\kappa_{\mathbf{n}}\left|\mathbf{r}_{A}-\boldsymbol{\rho}\right|}}{4\pi\left|\mathbf{r}_{A}-\boldsymbol{\rho}\right|}\phi_{\mathbf{n}}(\mathbf{r}_{B})\phi_{\mathbf{n}}(\boldsymbol{\rho})\right].
\label{eq:GreenInPosSpace}\end{multline}
The low-energy assumption (\ref{eq:ElasticScattering}) 
ensures that the only open exit channel corresponds to a $B$ particle in the ground vibrational state,
and therefore that
\begin{equation}
-\frac{\hbar^{2}\kappa_{\mathbf{n}}^{2}}{2m_{A}}=\frac{\hbar^{2}k^{2}}{2m_{A}}-\hbar\omega\left(n_{x}+n_{y}+n_{z}\right)<0,
\ \forall\ \mathbf{n}\neq0.
\end{equation}
As a consequence, the terms of Eq.~(\ref{eq:GreenInPosSpace}) involving
excited states of the harmonic oscillator give exponentially vanishing
contributions and can be neglected when $A$ is at a distance $\gg a_{\rm ho}$ from the center
of the harmonic well. Expanding $\left|\mathbf{r}_{A}-\boldsymbol{\rho}\right|\simeq r_{A}-
\boldsymbol{\rho}\cdot\mathbf{r}_{A}/r_{A}$ and substituting
into Eq.~(\ref{eq:ScatteringSolution}), we get
\begin{equation}
\psi(\mathbf{r}_{A},\mathbf{r}_{B})\stackrel{^{r_{A}\rightarrow\infty}}
{\simeq}\left[\frac{\sin(kr_{A})}{kr_{A}}+f_{k}\frac{\exp(ikr_{A})}{r_{A}}\right]\phi_{0}(r_{B}),\label{eq:Incoming+Outgoing}
\end{equation}
where we have introduced the scattering amplitude
\begin{equation}
f_{k}=-a\frac{m_{A}}{\mu}\int\textrm{d}\boldsymbol{\rho}
\,\frac{\sin(k\rho)}{k\rho}\phi_{0}(\rho)\psi_{\mathrm{reg}}(\rho).\label{eq:f_k}
\end{equation}
In the limit of zero-energy, the scattering amplitude defines the
\emph{effective scattering length} $a_{\textrm{eff}}$ 
through 
\begin{equation}
a_{\mathrm{eff}}\equiv
-\lim_{k\rightarrow0}f_{k}=a\frac{m_{A}}{\mu}\int\textrm{d}\boldsymbol{\rho}\,
\phi_{0}(\rho)\psi_{\mathrm{reg}}^{k=0}(\rho).\label{eq:a_eff}
\end{equation}

\begin{figure}
\begin{center}\includegraphics[ clip, width=\columnwidth]{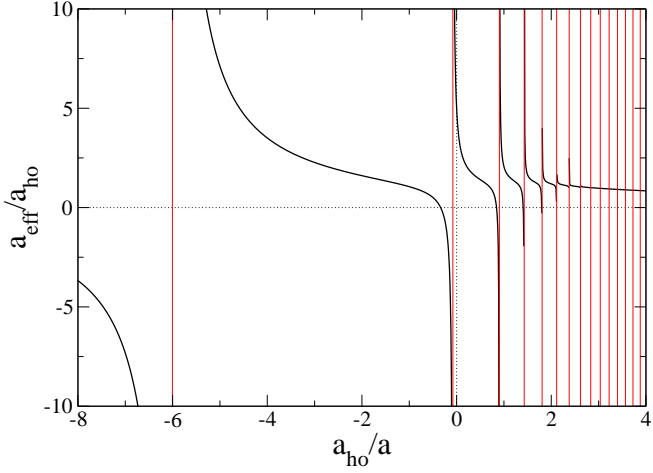}\end{center}

\caption{\label{cap:Effective0.15} (Color online) Effective scattering length $a_{\mathrm{eff}}$
(continuous line) as a function of $a^{-1}$ for $m_{B}/m_{A}=0.15$
(trapped $^{6}$Li and free $^{40}$K). The vertical lines (red) mark
the positions of the resonances (a step of $a_{\mathrm{ho}}/a=0.01$
is used to sample the curves, and some of the resonances are too narrow
to be seen on the graph).}
\end{figure}

\begin{figure}
\begin{center}\includegraphics[ clip, width=\columnwidth]{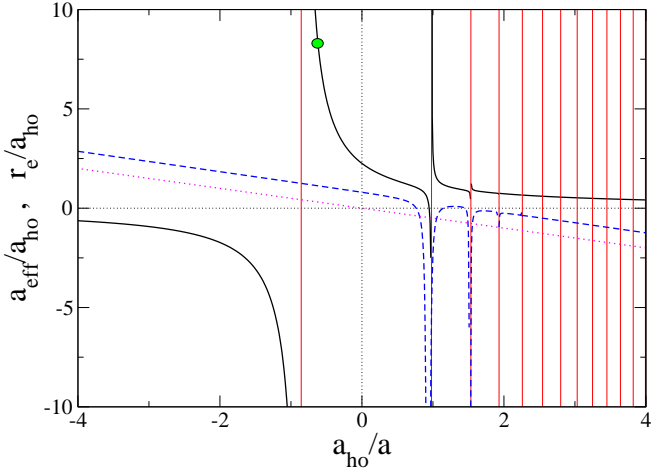}\end{center}

\caption{\label{cap:EffectiveEgales} (Color online) Same as Fig.~\ref{cap:Effective0.15}
for two particles of equal mass. The blue dashed line is the effective
range $r_{e}$, and the magenta dotted line is $r_{e,\mathrm{Born}}=-(\mu/m_A) a_{\rm ho}^2/a$,
obtained in the Born approximation by replacing 
$\psi_{\rm reg}(\rho)$ with $\psi_0(\rho,\rho)$
in Eq.~(\ref{eq:f_k}). A green dot at $a=-1.6a_{\textrm{ho}}$ marks
the position where $a_{\textrm{eff}}\approx d$ if $V_{0}^{B}=50E_{r}^{B}$.}
\end{figure}

\begin{figure}
\begin{center}\includegraphics[ clip, width=\columnwidth]{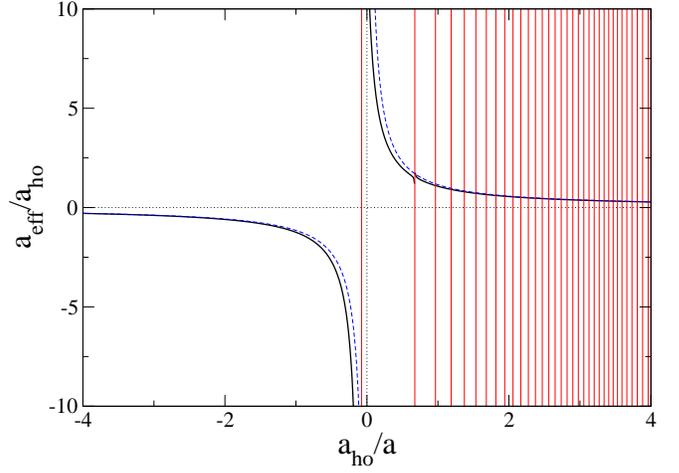}\end{center}

\caption{\label{cap:Effective6.67} (Color online) Same as Fig.~\ref{cap:Effective0.15}
for $m_{B}/m_{A}=6.67$ (trapped $^{40}$K and free $^{6}$Li). The
blue dashed line is the Born approximation $a_{\mathrm{eff,Born}}=a m_A/\mu$, obtained
by replacing $\psi_{\rm reg}^{k=0}(\rho)$ by $\phi_0(\rho)$ in Eq.~(\ref{eq:a_eff}).}
\end{figure}

We have solved numerically Eq.~(\ref{eq:IntegralEquation}), and in Figs.~\ref{cap:Effective0.15}-\ref{cap:Effective6.67}
we plot $a_{\mathrm{eff}}$ as a function of $1/a$ for different
values of $m_{B}/m_{A}$. We choose $m_{B}/m_{A}=0.15,1,6.67$, corresponding
to the physical cases of a mixture of $A={}^{40}$K and $B={}^{6}$Li,
a mixture of two different internal states of atoms of the
same species, and a mixture of $A={}^{6}$Li and $B={}^{40}$K,
respectively.  It is apparent that $a_{\rm eff}$ presents a series of intriguing resonances,
which can be used to tune $a_{\rm eff}$ to a high value and whose physical original are
discussed in \S\ref{subsec:reson}.

\begin{figure}
\begin{center}\includegraphics[ clip, width=\columnwidth]{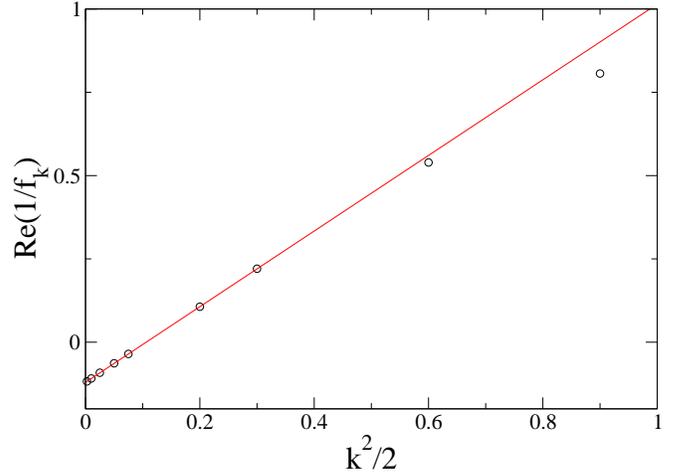}\end{center}

\caption{\label{cap:r_e} (Color online) Energy dependence of the scattering amplitude for
two particles of equal mass at $a=-1.6a_{\mathrm{ho}}$ (green marker
in Fig.~\ref{cap:EffectiveEgales}). The linear fit at low energy,
$Re(1/f_{k})=-1/a_{\mathrm{eff}}+r_{e}k^{2}/2$, yields $a_{\rm eff}=8.3
a_{\rm ho}$ and $r_{e}=1.135a_{\mathrm{ho}}$
(units of $a_{\mathrm{ho}}^{-2}$ and $a_{\mathrm{ho}}^{-1}$ on the
horizontal and vertical axis, respectively).}
\end{figure}

In Figs.~\ref{cap:EffectiveEgales} and \ref{cap:r_e}
we also plot, in the case $m_{B}=m_{A}$, the behavior of the effective range $r_e$.
It is defined as a coefficient in the low $k$ expansion of
the inverse scattering amplitude:
\begin{equation}
f_k^{-1}= -\left[a_{\rm eff}^{-1} + ik - r_e k^2/2 + \ldots\right].
\end{equation}
The replacement of a trapped $B$ particle by a fixed point-like scatterer, as done
in Eq.~(\ref{eq:Veff}), is allowed when the $k^2$ term in the above expansion is negligible,
i.e.\, when
\begin{equation}
|r_e| k^2 \ll |a_{\rm eff}^{-1}+ik|.
\label{eq:cond_contact}
\end{equation}
In the ideal regime for matter wave localization, $k< a_{\rm eff}^{-1}\sim 1/d$. Since
$r_{e}$ is generally of order $a_{\textrm{ho}}\ll d$, this condition Eq.~(\ref{eq:cond_contact}) is satisfied.
Note that $r_e$ diverges when $a_{\textrm{eff}}\rightarrow0$, a generic phenomenon: expanding $f_k$ rather than
$f_k^{-1}$ in powers of $k$,
\begin{equation}
f_k =- a_{\rm eff} + i k a_{\rm eff}^2 -\frac{1}{2} r_e a_{\rm eff}^2 k^2 +\ldots,
\end{equation}
one generically expects that $r_e a_{\rm eff}^2$ has a finite limit when $a_{\rm eff}\rightarrow 0$.
This was demonstrated analytically both for a square well \cite{Duine04} and a van der Waals interaction potential 
\cite{Flambaum99}.
In this limit, Eq.~(\ref{eq:cond_contact}) is thus violated for any finite $k$; however, this is not an ideal regime
to obtain matter wave localization.

\subsection{Resonances of the effective scattering length} \label{subsec:reson}

We now show that the effective coupling constant experiences an infinite set of resonances
due to the presence of the external confinement. 
In the limit $k\rightarrow 0$, the resolvent has real matrix elements in position space,
see Eq.~(\ref{eq:GreenInPosSpace});
so does the symmetric operator $\hat{O}$, which then admits real eigenvalues $\lambda_i$
and orthonormal eigenvectors $|e_i\rangle$ (see Eqs.(\ref{eq:UF1},\ref{eq:UF2})).
We then rewrite Eq.~(\ref{eq:a_eff}) as
\begin{align}
a_{\rm eff} &=a\frac{m_{A}}{\mu}\langle\tilde{\psi}_0^{k=0}|(\mbox{I}-g\,\hat O_{k=0})^{-1}|\tilde{\psi}_0^{k=0}\rangle \nonumber\\ 
    &=\frac{m_A}{2\pi\hbar^2} \sum_i \frac {|\langle\tilde{\psi}_0^{k=0}|\mathrm {e}_i \rangle |^2}{g^{-1}-\lambda_i}, 
\label{eq:f_k__rewritten}
\end{align}
where the scalar product of functions of the single variable $R$ is defined
as $\langle u|v\rangle = 4\pi \int_0^{+\infty} dR\, u^*(R) v(R)$.
This means that a singularity in
$a_{\mathrm{eff}}$ is expected whenever the denominator of Eq.~(\ref{eq:f_k__rewritten})
vanishes, i.e.~whenever the inverse free space scattering length $1/a$ equals
$2\pi\hbar^{2}\lambda_{i}/\mu$, 
provided that the non-interacting wave function has non-zero overlap with the corresponding eigenstate $|\mathrm {e}_i\rangle$. 

Confinement-induced resonances have been been analyzed
theoretically in different contexts, such as 1D wave guides \cite{Olshanii98,Bergeman03,Peano05},
3D optical lattices \cite{Fedichev04}, and quasi-2D condensates
\cite{Petrov00}. In 1D wave guides the effect is particularly remarkable:
due to the presence of the transverse confinement, a contact potential
acquires a bound state for any value of the 3D scattering length (while
in free space the contact potential has a bound
state only for $a>0$). Confinement-induced modifications of two-body
scattering properties have very recently been observed experimentally
by the Zurich group in 1D waveguides \cite{Moritz05} and in 3D optical
lattices \cite{Stoeferle05}. In most of the cited papers, the underlying
translational symmetry and the harmonic nature of the confinement
permit the factorization of the center-of-mass motion: this in
turn implies that a single confinement-induced resonance can exist, since only
one state in the closed channel is coupled to the open channel \cite{Bergeman03}.
When this factorization is not possible, as in our setup or in the
case of a 1D wave guide with anharmonic transverse confinement \cite{Peano05},
an infinite set of states in the closed channel has non-zero coupling
to the open one, and an infinite number of resonances appears. In
practice however only a few of them may be resolved and relevant in
an experiment since they become increasingly sharper as $a_{\textrm{ho}}/a$
becomes larger  (i.e.~$|\langle\tilde{\psi_0}|\mathrm {e}_i \rangle |^2\rightarrow 0$ for large $i$).
The position of these resonances can be predicted analytically in various limits,
depending on the sign of $a$, as we now discuss.

\subsubsection{Position of the resonances for $a>0$}

When $a>0$, the pseudopotential admits a bound state in which the
two particles can {}``sit'' for a variable time, forming a molecule
that oscillates in the harmonic well. To understand this point, we
rewrite the two-body Hamiltonian (\ref{eq:twobodyH}) as\begin{multline}
-\frac{\hbar^{2}\Delta_{R}}{2(m_{A}+m_{B})}+\frac{1}{2}m_{B}\omega^{2}R^{2}-\frac{\hbar^{2}\Delta_{r}}{2\mu}+g\cdot\delta(\mathbf{r})\cdot\frac{\partial}{\partial r}\left(r\,\cdot\right)\\
+\left[\frac{1}{2}\frac{m_{A}\mu}{m_A+m_B}
\omega^{2}r^{2}-\mu\omega^{2}\mathbf{R}\cdot\mathbf{r}\right],\label{eq:H_in_COM_and_rel_coord}\end{multline}
and treat the terms in the square parenthesis as a perturbation.
The unperturbed part admits the factorized eigenstates $\psi_{n,l=0}(\mathbf{R},\mathbf{r})=\phi_{n,l=0}(\mathbf{R})\chi(r)$
that describe a bound molecule with internal wavefunction
$\chi(r)=\exp(-r/a)/\sqrt{2\pi a}r$
and center-of-mass in an eigenstate of the harmonic oscillator of angular momentum $l=0$
and radial quantum number $n\geq 0$. Since
both the initial state and the Hamiltonian are spherically symmetric,
conservation of angular momentum allows only $l=0$ intermediate molecular states.
Within this unperturbed approximation, $a_{\mathrm{eff}}$ diverges each
time the energy of the oscillating molecule corresponds to the ground
state energy of the pair of atoms, i.e.~at the values of $a=a_{\mathrm{res}}$
that satisfy \begin{equation}
\left(2n+\frac{3}{2}\right)\hbar\omega\sqrt{\frac{m_{B}}{m_{A}+m_{B}}}
-\frac{\hbar^{2}}{2\mu a_{\mathrm{res}}^{2}}=\frac{3}{2}\hbar\omega.\label{eq:theoResPos}\end{equation}
As can be seen in the upper part of Fig.~\ref{cap:posRes}, this
formula describes the position of the resonances with $a>0$ in a
wide region of the graph, since corrections to it are only $O(a/a_{\mathrm{ho}})^{2}$ \cite{why_a2}.

\subsubsection{Existence of a resonance for $a<0$}

For $a<0$ and for a large enough
$m_A/m_B$ ratio, the presence of at least one resonance
can be demonstrated by a very simple variational argument performed at the unitary
limit $1/a=0$.
A contact potential characterized by the scattering length $a$ can be
replaced by the simple boundary condition $\psi(\mathbf{R},\mathbf{r})\stackrel{{r\rightarrow0}}{=}C(\mathbf{R})\left(r^{-1}-a^{-1}\right)+o(1)$,
where $C$ is an arbitrary function of the center of mass coordinate
$\mathbf{R}$. The $s$-wave Ansatz \begin{equation}
\psi(R,r)=\mathcal{N}\exp\left(-R^{2}/\lambda^{2}\right)\frac{\exp\left(-r^{2}/\sigma^{2}\right)}{r},\end{equation}
with $\lambda$ and $\sigma$ variational parameters, satisfies the
boundary condition imposed by a unitarity limited contact potential,
i.e.~characterized by $1/a=0$. Its energy can be calculated from
the Hamiltonian (\ref{eq:H_in_COM_and_rel_coord}):
the term $\mathbf{R\cdot r}$ has a vanishing contribution when
averaged over this state, and the variational energy assumes the minimum value
\begin{equation}
E^{\infty}=\frac{3}{2}\hbar\omega\sqrt{\frac{m_{B}}{m_{A}+m_{B}}}+\frac{1}{2}\hbar\omega\sqrt{\frac{m_{A}}{m_A+m_B}}.\end{equation}
We might imagine continuously tuning $a$ from $0^{-}$, where no
bound state can exist, towards $-\infty$. If $E^{\infty}<3\hbar\omega/2$
(i.e.~if $m_{A}/m_{B}>9/16$) a bound state for $1/a=0$ is guaranteed
by the former Ansatz, and at least one resonance for $a_{\mathrm{eff}}$
must exist in the $a<0$ region.

The following more elaborate calculation allows to prove the existence of a resonance for $a<0$, for an arbitrary
mass ratio $m_A/m_B$. One performs a variational calculation directly on the integral operator $\hat{O}^{k=0}$,
taking as a variational function $f(R)=R\exp(-R^2/2a_{\rm ho}^2)$.
Using Eqs.(\ref{eq:UF1},\ref{eq:UF2}) and performing Gaussian integrals we obtain in harmonic oscillator units:
\begin{eqnarray}
\langle f|\hat{O}^{k=0}|f\rangle &=& \left(\frac{\alpha}{2}\right)^{3/2} \int_0^{+\infty} d\tau\,
\left\{\left[(1+\alpha)\tau\right]^{-3/2} \right.\nonumber \\
&&\left.-\left[\tau+\alpha-\alpha \exp(-\tau)\right]^{-3/2}\right\}.
\end{eqnarray}
 From the inequality $\exp(-\tau)> 1-\tau$ valid for any $\tau>0$, we conclude that the integrand is a negative
function, so that $\langle f|\hat{O}^{k=0}|f\rangle <0$. Since the trial wavefunction $f(R)$ is
proportional to $\tilde{\psi}_0^{k=0}(R)=R\, \phi_0(R)$, we conclude that $\hat{O}^{k=0}$ admits
at least one eigenvector $|\mathrm{e}_i\rangle$ with a non-zero overlap with $\tilde{\psi}^{k=0}$ and with a negative
eigenvalue $\lambda_i<0$. 
The identity Eq.~(\ref{eq:f_k__rewritten}) then implies the existence of a resonance in $a_{\rm eff}$ for $a<0$.

\subsubsection{Position of the resonances for $a<0$}

\begin{figure}
\begin{center}\includegraphics[ clip, width=\columnwidth]{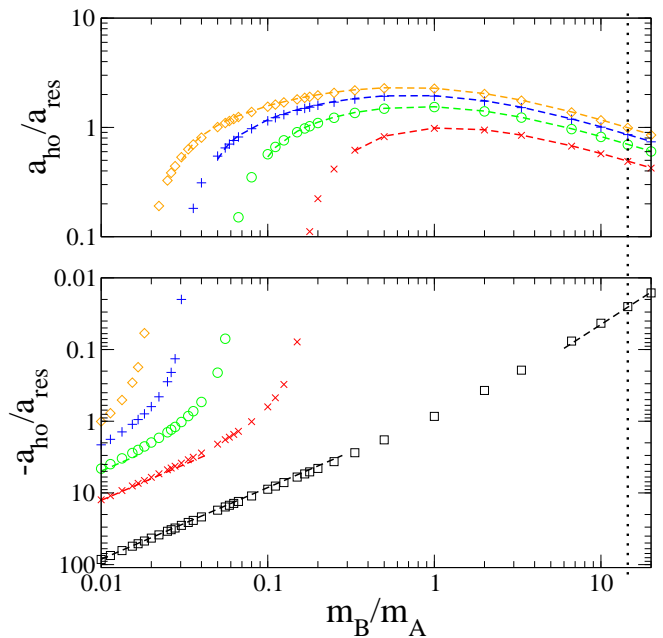}\end{center}

\caption{\label{cap:posRes}(Color online) Position of the broadest resonances for positive
(upper graph) and negative (lower graph) $a$ as a function of $m_{B}/m_{A}$.
The dashed lines are the theoretical predictions: for $a>0$ they are given by Eq.~(\ref{eq:theoResPos});
for $a<0$ and $m_B<m_A$ they are given by the Hamiltonian in Eq.~(\ref{eq:HubbardLikeHamilt});
for $a<0$ and $m_B>m_A$ it is given by Eq.~(\ref{eq:pos_asympt}).
In the upper graph, from top to bottom symbols correspond
to $n=4,3,2,1$ in Eq.~(\ref{eq:theoResPos}). The dotted vertical line indicates the value
of the mass ratio $m_B/m_A$ when $A={}^6$Li and $B={}^{87}$Rb.
}
\end{figure}

When $a<0$ and $m_{B}/m_{A}\ll1$, the position of the resonances
can be found 
with the aid of the Born-Oppenheimer approximation.
For a fixed position of the massive particle $A$, one calculates the energy
of the $B$ particle, which then constitutes an effective potential for
the $A$ particle.
Restricting for simplicity to the mean field regime $|a|\ll a_{\rm ho}$,
one can assume that the particle $B$ remains in the ground state of the well,
thereby creating an effective Gaussian attractive
well for the $A$ particle. Hence the effective Hamiltonian for $A$:
\begin{equation}
H_{\mathrm{eff}}=-\frac{\hbar^2}{2m_{A}}\Delta_{\mathbf{r}_A}
-\frac{2\pi\hbar|a|}{m_{B}}\frac{\exp(-r_{A}^{2}/a_{ho}^{2})}
{\left(\sqrt{\pi}a_{\mathrm{ho}}\right)^{3}}.\label{eq:HubbardLikeHamilt}
\end{equation}
As argued above, the characteristic range of the potential is of order
$a_{\textrm{ho}}$. This Hamiltonian can be easily solved numerically, and predicts
a divergence of $a_{\mathrm{eff}}$ whenever the combination $(|a|/a_{\mathrm{ho}})(m_{A}/m_{B})$
equals the critical value for the appearance of a new bound state
(see Fig.~\ref{cap:posRes}, dashed lines in lower graph). 

In the opposite limit $m_B/m_A \gg 1$, there is one resonance left on the $a<0$ side.
It is intuitive that its position $a_{\rm res}$ tends to $-\infty$ in this limit, the $B$
particle being then perceived by $A$ as a fixed scatterer of scattering length $a$, for which the resonance
is obtained for $a=-\infty$.
At $k=0$, one then expects that $\psi_{\rm reg}(R)\propto \phi_0(R)$.
To formalize this intuition, we expand the integral operator $\hat{O}$ in powers
of the mass ratio $\alpha=m_A/m_B\rightarrow 0$: from Eqs.(\ref{eq:UF1},\ref{eq:UF2}), we get
\begin{equation}
\hat{O}^{k=0} = \alpha^{3/2} \hat{O}_0 + \alpha^{5/2} \hat{O}_1+ \ldots
\end{equation}
It is then possible to check analytically, by calculation of Gaussian integrals,
that one has exactly $\hat{O}_0 |\tilde{\phi}_0\rangle =0$, with $\tilde{\phi}_0(R)=R\,\phi_0(R)$.
Using perturbation theory, we obtain the series expansion of the lowest eigenvalue
of $\hat{O}^{k=0}$:
\begin{equation}
\lambda_0 = \alpha^{5/2} \langle \tilde{\phi}_0| \hat{O}_1 |\tilde{\phi}_0\rangle
+ O(\alpha^{7/2}) =
-\frac{\alpha^{5/2}}{\pi\sqrt{2}}+ O(\alpha^{7/2})
\end{equation}
in harmonic oscillator units.
The resulting lowest order expression for the resonance position is
\begin{equation}
\frac{a_{\rm ho}}{a_{\rm res}} = -\sqrt{2} \left(\frac{m_A}{m_B}\right)^{3/2}\left[1+O\left(\frac{m_A}{m_B}\right)\right],
\label{eq:pos_asympt}
\end{equation}
and is shown as a dashed line in Fig.\ref{cap:posRes} \cite{better}.

\section{Experimental outlook and conclusions\label{sec:Conclusions}}

In this paper, we have shown numerically that the way of producing a disordered
potential for matter waves proposed in \cite{Gavish05}, i.e.\ the use of atoms
randomly trapped at the nodes of an optical lattice, indeed leads to the appearance
of exponentially localized states in a three-dimensional geometry.
Our numerical method allows to directly compute the wavefunction of the localized
states; it is based on the fact that the matrix elements of the resolvent
of the Hamiltonian can be calculated extremely efficiently for the interaction
of the matter wave with point scatterers, a fact already used with success in the context
of light localization \cite{Rusek95,Rusek00,Mandonnet}. The method also allows to obtain analytical results in
a straightforward manner, such as a check of the existence of localized states in 1D
from Furstenberg theorem or the derivation of a link between the inverse lifetime
of a localized state and its spatial decay in 3D, for a finite spatial extension of the 
scattering medium.

The main physical result is that numerous long-lived localized states appear
for a wave number of the matter wave smaller than the inverse of the mean separation between
scatterers, 
when the effective scattering length $a_{\rm eff}$ of the matter wave on a trapped atom
is positive and of the order of the mean separation of the scatterers.
For the numerical examples of this paper, with a 10\% occupancy of the lattice
sites and for a wavenumber of the matter wave as large
as $\sim 0.5/d$, there are localized states with localization lengths that can be as small as the lattice period $d$,
usually a sub-micron quantity. 
This extremely strong localization
allows to have very long-lived localized states (lifetime larger that
$10^4$ inverse recoil angular frequency of the matter wave)
even in disordered samples
with a radius as small as 10 lattice periods.
By a full solution of the scattering problem of a free atom with a harmonically
trapped one,
we have shown that the required large values of $a_{\rm eff}\sim d$ 
can be obtained by using an inter-species Feshbach resonance, and we have characterized intriguing
confinement-induced resonances that appear in this two-body scattering process.

How to proceed in a real experiment to get evidence of these localized states~?
A possibility is to extend to matter waves what was proposed
for light in \cite{Mandonnet}: one introduces 
the matter wave wavepackets inside the scattering medium
at a low value of $a_{\rm eff}$, then one tunes $a_{\rm eff}$ to the desired high value and
one lets the matter wave evolve in presence of the scattering medium (but
in the absence of an external trapping potential). After an adjustable time $\tau$, one
measures the number of remaining matter wave atoms $N_{\rm rem}(\tau)$ in the scattering
medium. Since the component of the matter wave wavefunction in localized states
decays exponentially in time with very weak rates $\Gamma$, the function
$N_{\rm rem}(\tau)$ should have a long tail, as compared to the case of a purely
ballistic or even diffusive expansion \cite{Wigner_time}. A further check that this long tail
has a decay rate varying exponentially with the size of the scattering medium 
would be a very convincing evidence of strong localization \cite{discrim}.

\begin{acknowledgments}
We are grateful to Uri Gavish, Dominique Delande, Martin Weitz, Leonardo Fallani,
Chris.~J.~Pethick, Henrik Smith, Anders S.~S\o rensen, Fabrice Gerbier,
Jean Dalibard and Christophe Salomon
for stimulating discussions. 
Laboratoire Kastler Brossel is a research unit of \'Ecole normale
sup\'erieure and of Universit\'e Pierre et Marie Curie, associated to
CNRS. Our research group in Paris is a member of the IFRAF Institute.
\end{acknowledgments}

\appendix
\section{Spatial decay of $\langle\mathbf{r}|{\mathcal G}(E+i0^+)|\mathbf{r'}\rangle$}
\label{appen:kinnon}

\begin{figure}
\begin{center}\includegraphics[ clip, width=\columnwidth]{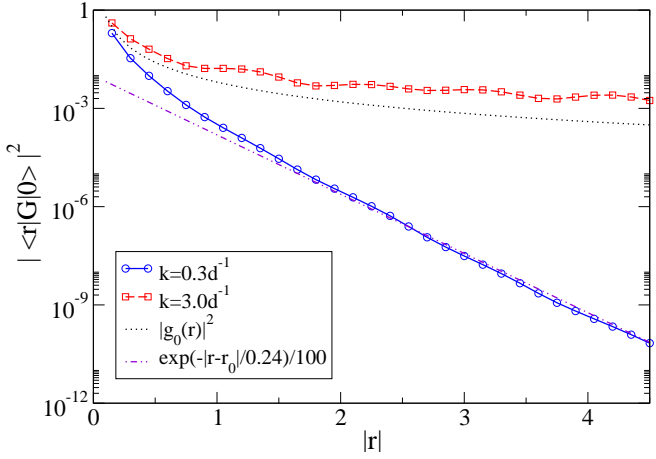}\end{center}

\caption{\label{cap:DecayingResolvent}(Color online) Decay of the 
off-diagonal real space matrix elements of the resolvent 
 as a function of the distance from the center of the cloud of scatterers (in units
of the lattice spacing $d$),
averaged over 100 different realizations of the random potential.
The $\approx4600$ scatterers ($a_{\textrm{eff}}=0.3d$) are distributed
in a cubic box of side $21d$, occupying each site of the lattice
with $p=0.5$. The y-axis is in units of $(\hbar^2/2m_A)^{-2}$.}
\end{figure}

As illustrated in Fig.~\ref{cap:DecayingResolvent}, in our model the off-diagonal matrix
elements of the resolvent $\langle\mathbf{r}|{\mathcal G}(E+i0^+)|\mathbf{r'}\rangle$ decay exponentially 
in $|\mathbf{r}-\mathbf{r'}|$ at low energy $E$.
Is this a signature of localization~? No, because this decay may be due to the fact
that the energy $E$ is in a spectral gap of the system. And this occurs as well
in the forbidden bands of a periodic system.

To illustrate this statement for our system, we have enclosed our lattice of scatterers
in a box of side $L$, imposing periodic boundary conditions on the
walls of the box. This amounts to replacing Eq.~(\ref{eq:G0Free})
by the particle propagator satisfying the correct boundary conditions:\begin{equation}
g_{0}^{\textrm{Box}}(\mathbf{r})=\frac{2m}{\hbar^{2}L^{3}}\sum_{\mathbf{q}}\frac{\textrm{e}^{i\mathbf{q}\cdot\mathbf{r}}}{k^{2}-\mathbf{q}^{2}},\label{eq:G0Box}\end{equation}
with $\mathbf{q}=2\pi\mathbf{n}/L$ and $\mathbf{n\in\mathbb{Z}^{3}}$
(a triplet of integers). Indeed we found that, for the ensemble of
scatterers used in Fig.~\ref{cap:DecayingResolvent}, the ground
state of the system, once enclosed in a box of side $L=23d$ (slightly
larger than the scattering medium), is characterized by a wave number
$k_{\textrm{min}}=0.7202d^{-1}$. The exponential decay shown at $k=0.3d^{-1}$
by $\left|\left\langle \mathbf{r}\left|\mathcal{G}\right|\mathbf{r}^{\prime}\right\rangle \right|^{2}$
is therefore simply indicating that at such low energy no state can
exist deep inside the medium. In a scattering experiment, we might
imagine a plane wave coming from infinity that scatters on the trapped
$B$ atoms: if $k<k_{\textrm{min}}$ the incoming wave undergoes total
reflection, and inside the random medium only penetrates an evanescent
wave, that decays exponentially from the boundary of the medium towards
its interior. This example clearly points out that an exponential
decay of $\left|\left\langle \mathbf{r}\left|\mathcal{G}\right|\mathbf{r}^{\prime}\right\rangle \right|^{2}$
is not a sufficient criterion to prove localization in our system,
since it does not guarantee the existence of states deep inside the
random potential.

In the criterion of Kramer and MacKinnon, introduced for a solid, the energy is
taken equal to the Fermi energy, with the assumption that the density
of states does not vanish at the Fermi energy.
This criterion then turns out not to be practical in our case,
since it requires a diagonalization of the Hamiltonian in order to calculate the density
of states.

\section{Analytical results in 1D}
\label{appen:1D}

We consider a 1D free space geometry with $N$ Dirac scatterers located in distinct positions $x_1 < \ldots < x_N$.
A quantum particle of mass $m$ interacts with the $N$ scatterers, with a coupling constant $g$.
Using the transfer matrix formalism we calculate the resolvent ${\mathcal G}(z)$, for $z=E+ i\eta$, $E$ real and $\eta> 0$,
which will give access to stationary wavefunctions in the limit $\eta\rightarrow 0^+$. 
Taking for simplicity $x_0=0$ different from the positions $x_i$,
we use the fact that $\psi(x)\equiv \langle x | {\mathcal G}(z) |x_0=0\rangle$ solves Schr\"odinger's equation with a source term:
\begin{equation}
\left[z+\frac{\hbar^2}{2m} \frac{d^2}{dx^2}- g \sum_{i=1}^{N} \delta(x-x_i) \right]\psi(x)= \delta(x).
\label{eq:Schr}
\end{equation}
Over a position interval containing neither one of the $x_i$ nor $x_0=0$, $\psi(x)$ is an eigenstate of $d^2/dx^2$
so that, introducing the unique $k_c$ such that $z=\hbar^2 k_c^2/2m$ with $\mbox{Im}\, k_c > 0$, one has:
\begin{equation}
\begin{array}{ccc}
x \leq x_1 & : &  \psi(x) = A_0^+ e^{i k_c x} + A_0^- e^{-i k_c x} \\
x_j \leq x \leq x_{j+1} &:& \psi(x) = A_j^+ e^{i k_c x} + A_j^- e^{-i k_c x} \\
x_s \leq x \leq 0 &:& \psi(x)= A_s^+ e^{i k_c x} + A_s^- e^{-i k_c x} \\
0 \leq x \leq x_{s+1} &:& \psi(x)= B_s^+ e^{i k_c x} + B_s^- e^{-i k_c x} \\
x_{j'} \leq x \leq x_{j'+1} &:&
\psi(x) = B_{j'}^+ e^{i k_c x} + B_{j'}^- e^{-i k_c x} \\
x_N \leq x &:& \psi(x) = B_N^+ e^{i k_c x} + B_N^- e^{-i k_c x} \\
\end{array}
\nonumber
\end{equation}
where we assumed that the first $s$ scatterer positions $x_1,\ldots, x_s$ are $<0$ and the $N-s$ other ones
$x_{s+1}, \ldots, x_{N}$ are $>0$, and where $j$ runs from 1 to $s-1$ and
$j'$ from $s+1$ to $N-1$. All the unknown coefficients $A^{\pm}$ and
$B^{\pm}$ shall now be determined from boundary conditions obeyed by $\psi(x)$.

A first boundary condition is that $\psi(x)$ should not diverge exponentially in $x=\pm\infty$.
This imposes $A_0^+=0$ and $B_N^-=0$.

The other boundary conditions originate from the fact that each Dirac distribution 
in Eq.~(\ref{eq:Schr}) introduces a discontinuity of the first order derivative of $\psi(x)$, whereas $\psi(x)$ 
remains continuous.
Integrating Eq.~(\ref{eq:Schr}) over an infinitesimal interval around $x_j$ leads to
$\psi'(x_j^+) - \psi'(x_j^-) = 2mg  \, \psi(x_j) /\hbar^2$
with $\psi(x_j^+)=\psi(x_j^-)=\psi(x_j)$.
These two equations allow to connect the unknown coefficients on the right of $x_j$ to the ones on the left by a 
two-by-two transfer matrix. For $j=1,\ldots,s$ we thus have
\begin{equation}
\left( \begin{array}{c} A_{j}^+ \\ A_{j}^- \\ \end{array} \right)
= P_{j} \left( \begin{array}{c} A_{j-1}^+ \\ A_{j-1}^- \\ \end{array} \right),
\end{equation}
and for $j=s+1,\ldots, N$:
\begin{equation}
\left( \begin{array}{c} B_{j}^+ \\ B_{j}^- \\ \end{array} \right)
= P_{j} \left( \begin{array}{c} B_{j-1}^+ \\ B_{j-1}^- \\ \end{array} \right).
\end{equation}
A simple calculation leads to the explicit expression
\begin{equation}
P_j = \left(\begin{array}{cc}
1-i\gamma & -i \gamma\, e^{-2ik_c x_j} \\
i \gamma\, e^{2ik_c x_j} & 1+i\gamma \\
\end{array} \right)
\label{eq:Pi}
\end{equation}
with $\gamma = m g /(\hbar^2 k_c)$.
We then introduce the two following matrices, one associated to the first $s$ scatterers,
\begin{equation}
{\cal A} \equiv P_s \ldots P_1
\end{equation}
and the other one to the $N-s$ last scatterers:
\begin{equation}
{\cal B} \equiv P_{s+1}^{-1} \ldots P_{N}^{-1}.
\end{equation}
They allow to express the coefficients $A_s^{\pm}$ in terms of $A_0^-$ and the coefficients
$B_s^{\pm}$ in terms of $B_N^+$:
\begin{eqnarray}
\left(\begin{array}{c} A_s^+ \\ A_s^- \end{array}\right)&=& {\cal A} 
\left(\begin{array}{c} 0 \\ A_0^{-} \end{array} \right) \label{eq:gros1}\\
\left(\begin{array}{c} B_s^+ \\ B_s^- \end{array}\right)&=& {\cal B} 
\left(\begin{array}{c} B_N^+ \\ 0 \end{array} \right).
\label{eq:gros2}
\end{eqnarray}

The last two unknowns, $A_0^-$ and $B_N^+$, are obtained from the boundary conditions
imposed by $\delta(x)$ in Eq.~(\ref{eq:Schr}), $\psi(0^-)=\psi(0^+)$ and
$\psi'(0^+)-\psi'(0^-)= 2m/\hbar^2,$
which imposes two equations on the coefficients
$B_s^{\pm}, A_s^{\pm}$, 
\begin{eqnarray}
A_s^+ &=& B_s^+ -\frac{m}{i\hbar^2 k_c} \\
A_s^- &=& B_s^- +\frac{m}{i\hbar^2 k_c}.
\end{eqnarray} 
Combined with Eq.~(\ref{eq:gros1}) and Eq.~(\ref{eq:gros2}), this leads to the following system:
\begin{equation}
\left(\begin{array}{cc}
{\cal B}_{11} & -{\cal A}_{12}\\
{\cal B}_{21} & -{\cal A}_{22} \\
\end{array}
\right)
\left(\begin{array}{c}
B_N^+ \\ A_0^-  \\
\end{array}
\right)
=
\frac{m}{i\hbar^2 k_c} \left(\begin{array}{r}
1 \\ -1 \\
\end{array}
\right),
\end{equation}
which can be solved explicitly:
\begin{eqnarray}
A_0^- &=& \frac{m}{i\hbar^2 k_c} \, \frac{{\cal B}_{11}+{\cal B}_{21}}{{\cal B}_{11}{\cal A}_{22}-{\cal B}_{21}{\cal A}_{12}} \\
B_N^+ &=& \frac{m}{i\hbar^2 k_c} \, \frac{{\cal A}_{12}+{\cal A}_{22}}{{\cal B}_{11}{\cal A}_{22}-{\cal B}_{21}{\cal A}_{12}}.
\end{eqnarray}

We then proceed with the limit $z$ tending to a real and positive energy $E$: $\eta\rightarrow 0^+$ and 
$k_c \rightarrow k= (2mE)^{1/2}/\hbar$. The $P_j$ then become physically meaningful transfer matrices in the
$SU(1,1)$ group. In the main text we introduced the reflection and transmission coefficients $r_+$, $t_+$ of
the last $N-s$ first scatterers on the axis oriented from $x=-\infty$ to $x=+\infty$.
This means that there exists a stationary solution of the usual Schr\"odinger equation equal
to $e^{ikx}+r_+ e^{-ikx}$ for $x<x_{s+1}$ and equal to $t_+ e^{ikx}$ for $x>x_N$. By definition of the
transfer matrices:
\begin{equation}
{\cal B} \left(\begin{array}{c}t_+ \\ 0 \end{array}\right) = \left(\begin{array}{c} 1 \\ r_+\end{array}\right).
\end{equation}
Similarly, introducing the reflection and transmission coefficients $r_-,t_-$ of the first $s$ scatterers
on the axis oriented this time from $x=+\infty$ to $x=-\infty$, we imply the existence
of a solution of Schr\"odinger's equation equal to $e^{-ikx} + r_- e^{ikx}$ for $x>x_s$
and equal to $t_- e^{-ikx}$ for $x< x_1$, which imposes
\begin{equation}
{\cal A} \left(\begin{array}{c} 0 \\ t_- \end{array}\right) = \left(\begin{array}{c} r_- \\  1\end{array}\right).
\end{equation}
This allows a physical interpretation of the coefficients of ${\cal A}$ and ${\cal B}$:
\begin{eqnarray}
{\cal A}_{12} = r_-/t_-, && {\cal A}_{22} = 1/t_-  \\
{\cal B}_{21} = r_+/t_+, && {\cal B}_{11} = 1/t_+.
\end{eqnarray}
This leads to very simple expressions for the Green's function $\psi(x)$ out of the scattering medium
and over the interval between two scatterers containing $x_0=0$:
\begin{eqnarray}
x\leq x_1: \psi(x) &=& \frac{m}{i\hbar^2 k} \, \frac{t_- (1+r_+)}{1-r_+ r_-} e^{-ikx} \nonumber \\
x_{s}\leq x\leq 0: \psi(x) &=& \frac{m}{i\hbar^2 k} \, \frac{(1+r_+)\left(e^{-ikx}+r_-e^{ikx}\right)}{1-r_+ r_-} \nonumber \\
0\leq x\leq x_{s+1}: \psi(x) &=& \frac{m}{i\hbar^2 k} \, \frac{\left(e^{ikx}+r_+e^{-ikx}\right)(1+r_-)}{1-r_+ r_-} \nonumber \\
x\geq x_N: \psi(x) &=& \frac{m}{i\hbar^2 k} \, \frac{t_+ (1+r_-) }{1-r_+ r_-} e^{+ikx}. \nonumber
\end{eqnarray}

\section{An approximate relation for some expectation values of $\mbox{Im}\, M$}
\label{appen:magique}

This appendix is useful to derive Eq.~(\ref{eq:decay_to_m0}), an equation which
relates the spatial decay of $\phi(\mathbf{r};\mathbf{r}_0)$ to the smallness of an eigenvalue of $M$.

Let us assume that we are at a positive energy $E$ such that the matrix $M$ has an eigenvalue $m_0$
extremely close to zero, much closer to zero anyway than all the other eigenvalues.
Let $\vec{v_0}$ be the associated eigenvector of $M$. The corresponding adjoint vector is an eigenvector
of $M^\dagger$ with the eigenvalue $m_0^*$. Since the matrix $M$ is (complex) symmetric, $M^\dagger$ is
simply $M^*$, the complex conjugate of $M$, so that one may take as adjoint vector the complex conjugate
$\vec{v_0}^*$ of $\vec{v_0}$. The imposed normalization condition is then
\begin{equation}
\langle v_0^*|v_0\rangle = \vec{v_0}\,^2=1.
\label{eq:norm}
\end{equation}
For compactness we use here Dirac's notation,
even if $M$ does not act in a Hilbert space.  

One has to calculate the vector $\vec{d}$ to fully determine $\phi(\mathbf{r};\mathbf{r}_0)$,
see Eq.~(\ref{eq:phi_jolie}). This vector solves the linear system
\begin{equation}
M\vec{d} = \vec{s}
\end{equation}
with the source term $s_j=-\exp{(ik|\mathbf{r}_j-\mathbf{r}_0|)}/|\mathbf{r}_j-\mathbf{r}_0|$.
As the eigenvalue $m_0$ of $M$ is the only one to be extremely close to zero, we take the approximate
expression
\begin{equation}
M^{-1} \simeq \frac{1}{m_0} \, |v_0\rangle \langle v_0^*|.
\label{eq:approx_M}
\end{equation}
This leads to
\begin{equation}
\vec{d} \simeq \frac{1}{m_0} \vec{v_0} \left(\vec{v_0}\cdot \vec{s}\right).
\label{eq:approx_d}
\end{equation}

To have access to an estimate of $\phi_{\textrm{out}}$,
we have to calculate the expectation value of the imaginary part of the matrix $M$ 
on the vector $\vec{d}$.
Since $M$ is symmetric, both the real part and the imaginary part of $M$ are hermitian matrices, with real
expectation values, so that
\begin{equation}
\vec{d}\,^*\cdot (\mbox{Im}\, M)\vec{d} = \mbox{Im}\, (\vec{d}\,^* \cdot M \vec{d})
= \mbox{Im}\, (\vec{d}\,^*\cdot\vec{s}).
\label{eq:pour_appen_suiv}
\end{equation}
Then using the approximation Eq.~(\ref{eq:approx_d}) we obtain
\begin{equation}
\vec{d}\,^*\cdot (\mbox{Im}\, M)\vec{d} \simeq \mbox{Im}\, \left(\frac{(\vec{v_0}^*\cdot\vec{s}\,^*)(\vec{v_0}^*\cdot\vec{s}\,)}{m_0^*}\right).
\label{eq:approx_out}
\end{equation}
An immediate application of this result is that the quantity $f(\mathbf{n})= \sum_j d_j e^{-i k\mathbf{n}\cdot\mathbf{r}_j}$, where $\mathbf{n}$
is a unity vector,
is typically much larger than unity. The calculation of the average of $|f|^2$ over the unit sphere indeed leads to
\begin{equation}
\langle |f|^2\rangle_{\mathbf{n}} = \frac{\vec{d}\,^*\cdot (\mbox{Im}\, M)\vec{d}}{k}.
\end{equation}
Since $|f|\gg 1$ in the low $m_0$ limit \cite{careful}, it is correct to neglect
the term $e^{-i k \mathbf{n}\cdot\mathbf{r}_0}$ in Eq.~(\ref{eq:far_field}) as was done in the main text.

To have access to an estimate of $\phi_{\textrm{in}}$, we have to calculate the maximal value of 
all the $|\mbox{Im}\, d_j|$. Let us call $n$ the index such that $|\mbox{Im}\, d_n|$ is the biggest one. According to the approximation
Eq.~(\ref{eq:approx_d}), we then have
\begin{equation}
\mbox{Im}\, d_n \simeq \mbox{Im}\left[\frac{v_{0,n} (\vec{v_0}\cdot\vec{s})}{m_0}\right]
\label{eq:approx_in}
\end{equation}
where $v_{0,n}$ denotes the component $n$ of the vector $\vec{v_0}$.

The expressions Eqs.(\ref{eq:approx_out},\ref{eq:approx_in}) greatly simplify when one chooses a position $\mathbf{r}_0$
that tends towards $\mathbf{r}_n$. The fact that the results shall not depend on this specific choice of $\mathbf{r}_0$ is established 
in the appendix \ref{appen:math}.
In this limit, all the components $s_j$ of the source term $\vec{s}$ are negligible as compared to
$s_n \sim -1/|\mathbf{r}_0-\mathbf{r}_n|$ so that $\vec{v_0}\cdot\vec{s}\sim -v_{0,n}/|\mathbf{r}_0-\mathbf{r}_n|$ and so on.
Then Eq.~(\ref{eq:est_int}) reduces to
\begin{equation}
\frac{\phi_{\rm in}^2}{\phi_{\rm out}^2} 
\simeq \frac{k R^2}{l^2} \mbox{Im}\, \left[\frac{v_{0,n}^{*2}}{m_0^*}\right].
\label{eq:c9}
\end{equation}
As shown in the appendix \ref{appen:math}, the components of the vector $\vec{v_0}$ are real in the limit of a vanishing 
$|m_0|$ so that $v_{0,n}^2$ may be pulled out of the imaginary part. Since the $|v_{0,j}|^2$ decrease roughly exponentially
at large distances $|\mathbf{r}_j-\mathbf{r}_n|$ over a length scale $b$ of the order of the mean scatterer separation,
as was known from studies of light localization
\cite{Rusek95,Rusek00,Mandonnet}, the normalization condition Eq.~(\ref{eq:norm}) leads
to $v_{0,n}^2 \sim 1/(\rho b^3)\lesssim 1$, where $\rho$ is the mean scatterers density. We then
get Eq.~(\ref{eq:decay_to_m0}).

\section{On the fact that some quantities are almost real}
\label{appen:math}

We consider here a value of $k$ such that the matrix $M$ has one (and only one) eigenvalue $m_0$ of
extremely small modulus. We then show that the corresponding eigenvector $\vec{v_0}$ of $M$,
normalized as 
in Eq.~(\ref{eq:norm}), is close to a vector with real components, a property used
in the Appendix \ref{appen:magique} and necessary to obtain the equation Eq.~(\ref{eq:decay_to_m0}).

First we give a physical argument. Starting from the approximation Eq.~(\ref{eq:approx_M}), 
and keeping terms only to leading order in $1/m_0$, we obtain from Eq.~(\ref{eq:phi_jolie}):
\begin{equation}
\phi(\mathbf{r};\mathbf{r}_0) \simeq -A\, \mbox{Im}\, \left[\frac{F(\mathbf{r}) F(\mathbf{r}_0)}{m_0}\right]
\label{eq:useful_phi}
\end{equation}
where
\begin{equation}
F(\mathbf{r}) = \sum_{j=1}^{N} v_{0,j} \frac{e^{ik|\mathbf{r}-\mathbf{r}_j|}}{|\mathbf{r}-\mathbf{r}_j|}.
\end{equation}
Now let us assume that the phase of $F(\mathbf{r}_0)$ (modulo $\pi$) changes significantly (that is in a way not
tending to zero with $m_0$) when $\mathbf{r}_0$ is varied. This means that the $\mathbf{r}$ dependent
wavefunction $\phi(\mathbf{r};\mathbf{r}_0)$ spans a subspace of the Hilbert space of dimension 2 when
$\mathbf{r}_0$ is varied, a subspace generated by the wavefunctions $\mbox{Re}\, F(\mathbf{r})$
and $\mbox{Im}\, F(\mathbf{r})$, or equivalently by $F(\mathbf{r})$ and $F^*(\mathbf{r})$.
This leads to the physically suspect situation that {\it two} independent localized states can be associated to
a given resonance. 

To avoid this suspect situation, one is led to the assumption that the function $F(\mathbf{r})$ 
has a constant phase (modulo $\pi$).
Taking the limit $\mathbf{r}\rightarrow \mathbf{r}_j$, this implies that all the $v_{0,j}$ have the same
phase (modulo $\pi$). From the normalization condition Eq.~(\ref{eq:norm}) we conclude
that all the $v_{0,j}$ are real, apart from small terms tending to zero with $m_0$.
Furthermore, this leads to the conclusion that the function $F(\mathbf{r})$ has to be real everywhere, 
apart from terms that tend to zero with $m_0$ \cite{a_l_infini}. A consequence that is important for the appendix
\ref{appen:magique} is that
$\vec{v_0}^*\cdot\vec{s}\simeq \vec{v_0}^*\cdot\vec{s}\,^*= -F(\mathbf{r}_0)^*$ are approximately real,
so that the result Eq.~(\ref{eq:c9}) does not depend indeed on the specific choice of $\mathbf{r}_0$.

Now we give a mathematical argument to show that $\vec{v_0}$ is almost real. 
We start from $M \vec{v_0} = m_0 \vec{v_0}$ and we split
the matrix $M$ in a real part and an imaginary part, $M=M_R + i M_I$. Since $M$ is symmetric,
both $M_R$ and $M_I$ are real symmetric. Taking the squared norm of the identity
$M_R \vec{v_0} = (m_0 I-i M_I) \vec{v_0}$,
where $I$ is the $N\times N$ identity matrix, leads to
\begin{equation}
||M_R \vec{v_0}||^2 = \vec{v_0}^*\cdot \left[M_I^2 + \mbox{Re}\, (m_0^2) \, I\right]\vec{v_0},
\end{equation}
where we used the fact that 
\begin{equation}
\vec{v_0}^*\cdot M_I \vec{v_0}= \mbox{Im}\,(m_0) \vec{v_0}^*\cdot\vec{v_0},
\label{eq:sum_rule}
\end{equation}
that can be proved as in Eq.~(\ref{eq:pour_appen_suiv}).
Next, we expand $\vec{v_0}$ in the orthonormal eigenbasis of $M_I$,
$\vec{v_0}= \sum_\alpha c_\alpha |m_{I,\alpha}\rangle,$
so that
\begin{equation}
\vec{v_0}^*\cdot M_I^2 \vec{v_0}= \sum_\alpha |c_\alpha|^2 m_{I,\alpha}^2.
\label{eq:expansion}
\end{equation}
It remains to use the following property of $M_I$, valid for an arbitrary vector $\vec{x}$:
\begin{equation}
\vec{x}\,^*\cdot M_I \vec{x} = k\int \frac{d^2 \mathbf{n}}{4\pi} |\sum_j x_j e^{ik\mathbf{n}\cdot\mathbf{r}_j}|^2
\end{equation}
to show that $0\leq m_{I,\alpha}\leq Nk$ \cite{aide}. As a consequence, $m_{I,\alpha}^2\leq
Nk m_{I,\alpha}$. Using the expansion Eq.~(\ref{eq:expansion}) and 
Eq.~(\ref{eq:sum_rule}), one gets an upper bound on $\vec{v_0}^*\cdot
M_I^2 \vec{v_0}$ so that
\begin{equation}
||M_R \vec{v_0}||^2 \leq \left[Nk\mbox{Im}\, m_0 + \mbox{Re}\,(m_0^2)\right] ||\vec{v_0}||^2.
\end{equation}

In the large scattering medium diameter $L\rightarrow +\infty$, an exponential decrease of the imaginary
part of $m_0$ is expected, $\sim \exp(-L/\xi)$. Close to the center of a resonance peak, $|\mbox{Re}\, m_0|
\sim\mbox{Im}\, m_0$, so that $||M_R \vec{v_0}||/||\vec{v_0}||= O\left[Nk \exp(-L/2\xi)\right]$.
On the contrary, the density of states for the spectrum of $M_R$ is not expected to be exponentially peaked:
the spacing between successive eigenvalues is expected to scale as $1/N$ at most. This is apparent
in Fig.\ref{cap:Poles-of-the-resolvent}, and we have checked it numerically for increasing
numbers of scattering centers. This implies
that $\vec{v_0}$, when expanded on the eigenvectors of $M_R$, populates essentially one eigenvector
of $M_R$, the one with the eigenvalue nearest to zero. Since this (unit norm) eigenvector of $M_R$ is proportional
to a real eigenvector, we deduce from Eq.~(\ref{eq:norm})  that the components of $\vec{v_0}$ are almost real, at the
$O(\mbox{Im}\, m_0)^{1/2}$ accuracy level.

\section{Integral equation for $\psi_{\mathrm{reg}}$ 
\label{appen:Inte}}

To calculate the non-interacting two-particle Green's function numerically,
we adapt a technique used in \cite{BCS_Yvan}.
The Feynman propagator $K$ associated with the non-interacting Hamiltonian
$H_{0}$ \cite{FeynmanBook},\begin{equation}
K_{t}(\mathbf{r}_{A},\mathbf{r}_{B};\boldsymbol{\rho}_{A},\boldsymbol{\rho}_{B})=\left\langle \mathbf{r}_{A},\mathbf{r}_{B}\left|e^{-iH_{0}t/\hbar}\right|\boldsymbol{\rho}_{A},\boldsymbol{\rho}_{B}\right\rangle \end{equation}
 can be factorized as the product of the two factors, $K_{t}=K_{t}^{A}K_{t}^{B}$,
the first term describing a free particle,\begin{equation}
K_{t}^{A}(\mathbf{r}_{A};\boldsymbol{\rho}_{A})=e^{-i\pi3/4}\left(\frac{m_{A}}{2\pi\hbar t}\right)^{3/2}\exp\left(\frac{im_{A}}{2\hbar}\frac{|\mathbf{r}_{A}-\boldsymbol{\rho}_{A}|^{2}}{t}\right)\label{eq:FreeProp}\end{equation}
and the second one a particle in a 3D harmonic oscillator,\begin{multline}
K_{t}^{B}(\mathbf{r}_{B};\boldsymbol{\rho}_{B})=\phi(t)e^{-i\pi3/4}\left(\frac{m_{B}\omega}{2\pi\hbar|\sin(\omega t)|}\right)^{3/2}\\
\exp\left(\frac{im_{B}\omega}{\hbar}\left[\frac{r_{B}^{2}+\rho_{B}^{2}}{2\tan(\omega t)}-\frac{\mathbf{r}_{B}\cdot\boldsymbol{\rho}_{B}}{\sin(\omega t)}\right]\right)\label{eq:HOProp}\end{multline}
where $\phi(t)=\exp(i\pi n/2)$ for $n\pi<\omega t<(n+1)\pi$. Setting
$K_{t}=0$ for $t<0$, $G_{E}$ is obtained
as the Fourier transform of $K_t$,\begin{gather}
G_{E}=-\frac{i}{\hbar}\int_{0}^{\infty}dt\, e^{i(E+i0^{+})t/\hbar}K_{t}.\label{eq:GasFourierOfK}\end{gather}

For simplicity, we introduce dimensionless variables by expressing
quantities in harmonic oscillator units and, even though the derivation
has been carried out for a generic mass ratio, we restrict ourselves
here to the special case $m_{A}=m_{B}=m$ 
(see end of appendix for the case $k\rightarrow 0$
with an arbitrary mass ratio $m_A/m_B$).

In order to find the equation satisfied by $\psi_{\mathrm{reg}}$,
we rewrite Eq.~(\ref{eq:ScatteringSolution}) as\begin{multline}
\psi(\mathbf{r}_{A},\mathbf{r}_{B})=\psi_{0}(\mathbf{r}_{A},\mathbf{r}_{B})+g\psi_{\mathrm{reg}}(R)\int d\boldsymbol{\rho}\, G_{E}\left(\mathbf{r}_{A},\mathbf{r}_{B};\boldsymbol{\rho},\boldsymbol{\rho}\right)\\
+g\int d\boldsymbol{\rho}\, G_{E}\left(\mathbf{r}_{A},\mathbf{r}_{B};\boldsymbol{\rho},\boldsymbol{\rho}\right)\left[\psi_{\mathrm{reg}}(\rho)-\psi_{\mathrm{reg}}(R)\right]\label{eq:ScatteringSolutionRewritten}\end{multline}
and let the two particles approach each other. We introduce here the
center-of-mass and relative coordinates $\mathbf{R}=(\mathbf{r}_{A}+\mathbf{r}_{B})/2$
and $\mathbf{r}=\mathbf{r}_{A}-\mathbf{r}_{B}$. Let us turn our attention
to the first integral appearing in Eq.~(\ref{eq:ScatteringSolutionRewritten}):\begin{equation}
U\equiv\int d\boldsymbol{\rho}\, G_{E}\left(\mathbf{R}+\frac{\mathbf{r}}{2},\mathbf{R}-\frac{\mathbf{r}}{2};\boldsymbol{\rho},\boldsymbol{\rho}\right).\end{equation}
Performing the Gaussian integration over $\boldsymbol{\rho}$ \cite{gaussInt}
one finds \begin{align}
U & =\frac{e^{i\pi3/4}}{\left(2\pi\right)^{3/2}}\int_{0}^{\infty}dt\,\frac{\tilde{\phi}(t)e^{i(Et+W)}}{|\sin(t)+t\cos(t)|^{3/2}},\label{eq:U}\end{align}
where \begin{equation}
W=\frac{|\mathbf{R}+\frac{\mathbf{r}}{2}|^{2}}{2t}+\frac{\left|\mathbf{R}-\frac{\mathbf{r}}{2}\right|^{2}}{2\tan(t)}-\frac{1}{x}\left|\frac{\mathbf{R}+\frac{\mathbf{r}}{2}}{2t}+\frac{\mathbf{R}-\frac{\mathbf{r}}{2}}{2\sin(t)}\right|^{2}.\end{equation}
 We have here introduced the shorthand notation $x(t)=[1/t+1/\tan(t)]/2$
and the phase factor $\tilde{\phi}(t)$, which equals $\exp(i\pi n/2)$
for $t_{n}<t<t_{n+1}$, where $t_{0}=0$ and $t_{1}=2.029,\, t_{2}=4.913,\,\dots$
are the consecutive solutions of $x(t)=0$. In the latter expression,
Eq.~(\ref{eq:U}), the contribution of the neighborhood
of $t=0$ diverges as $1/r$ for $r\rightarrow0$:
\begin{alignat}{1}
\frac{\left(2\pi\right)^{3/2}}{e^{i\pi3/4}}U & =\int_{0}^{\delta}dt\,\frac{e^{ir^{2}/4t}}{(2t)^{3/2}}+\int_{\delta}^{\infty}dt\,\dots\nonumber \\
 & =\int_{0}^{\infty}dt\,\frac{e^{ir^{2}/4t}}{(2t)^{3/2}}+\int_{\delta}^{\infty}dt\,\left(-\frac{1}{(2t)^{3/2}}+\dots\right)\nonumber \\
 & =\frac{1}{r}\sqrt{\frac{\pi}{2}}e^{i\pi/4}+\int_{\delta}^{\infty}dt\,\left(-\frac{1}{(2t)^{3/2}}+\dots\right).\end{alignat}
We finally find\begin{multline}
g\psi_{\mathrm{reg}}(R)\int d\boldsymbol{\rho}\, G_{E}\left(\mathbf{R}+\frac{\mathbf{r}}{2},\mathbf{R}-\frac{\mathbf{r}}{2};\boldsymbol{\rho},\boldsymbol{\rho}\right)\\
\stackrel{r\rightarrow0}{=}\psi_{\mathrm{reg}}(R)\left(gF_{1}(R)-\frac{a}{r}\right)+o(1)\end{multline}
that separates out the expected $1/r$ divergent contribution and
defines \begin{equation}
F_{1}(R)=\frac{e^{i\pi3/4}}{\left(2\pi\right)^{3/2}}\int_{0}^{\infty}\textrm{d}t\,\left(\frac{\tilde{\phi}(t)e^{i(Et+W_{0})}}{|\sin(t)+t\cos(t)|^{3/2}}-\frac{1}{(2t)^{3/2}}\right)\end{equation}
with $W_{0}=R^{2}\left(\cos(t)-t\sin(t)/2-1\right)/\left(\sin(t)+t\cos(t)\right)$.
The term $(2t)^{-3/2}$ regularizes $F_{1}$ in the neighborhood of
$t=0$, and we have taken the limit $\delta\rightarrow0$.

Let us now consider the remaining term appearing in Eq.~(\ref{eq:ScatteringSolutionRewritten}):
when $\mathbf{r}\rightarrow0$ and $\mathbf{u}=\mathbf{R}-\boldsymbol{\rho}\rightarrow0$,
the Green's function $G_{E}$ diverges as $|\mathbf{u}|^{-4}$, or
equivalently as $\left(|\mathbf{r}_{A}-\boldsymbol{\rho}_{A}|^{2}+|\mathbf{r}_{B}-\boldsymbol{\rho}_{B}|^{2}\right)^{-2}$,
but the second integral in Eq.~(\ref{eq:ScatteringSolutionRewritten})
is convergent in $\mathbf{u}\sim0$: \begin{multline}
\int d\boldsymbol{\rho}\, G_{E}\left(\mathbf{R}+\frac{\mathbf{r}}{2},\mathbf{R}-\frac{\mathbf{r}}{2};\boldsymbol{\rho},\boldsymbol{\rho}\right)\left[\psi_{\mathrm{reg}}(\rho)-\psi_{\mathrm{reg}}(R)\right]\\
\stackrel{\mathbf{r}\rightarrow0}{=}\int\frac{d\mathbf{u}}{\left|\mathbf{u}\right|^{4}}\left[\mathbf{u}\left.\frac{\partial\psi_{reg}}{\partial\mathbf{u}}\right|_{\mathbf{u=0}}+O\left(\left|\mathbf{u}\right|^{2}\right)\right]\end{multline}
 (the first order term vanishes due to spherical symmetry). In this
term therefore no divergence arises and we may set $r=0$. The angular
integrations can be performed analytically and we obtain:
\begin{multline}
\int d\boldsymbol{\rho}\, G_{E}\left(\mathbf{R},\mathbf{R};
\boldsymbol{\rho},\boldsymbol{\rho}\right)\left[\psi_{\mathrm{reg}}(\rho)-\psi_{\mathrm{reg}}(R)\right]\\
=\frac{1}{R}\int_{0}^{\infty}d\rho\,\left[\tilde{\psi}_{\mathrm{reg}}(\rho)-\frac{\rho}{R}\tilde{\psi}_{\mathrm{reg}}(R)\right]F_{2}(R,\rho)\end{multline}
with\begin{equation}
F_{2}(R,\rho)=\int_{0}^{\infty}dt\,\frac{\phi(t)e^{i[Et+(R^{2}+\rho^{2})x]}}{\left(t|\sin(t)|\right)^{3/2}}\frac{\sin(2R\rho y)}{\left(2\pi\right)^{2}y},\end{equation}
 which is symmetric under the exchange of $\rho$ and $R$. We have
here introduced the radial wave function $\tilde{\psi}(R)=R\psi(R)$
and the function $y(t)=[1/t+1/\sin(t)]/2$.

Writing \begin{equation}
\psi\left(\mathbf{r}_{A},\mathbf{r}_{B}\right)\stackrel{r\rightarrow0}{=}\psi_{\mathrm{reg}}(R)\left(1-\frac{a}{r}\right)+o(1),\label{eq:PsiCloseTo0}\end{equation}
 we can cancel the divergent contribution on both sides of Eq.~(\ref{eq:ScatteringSolutionRewritten}),
and the remaining terms constitute the implicit integral equation
\begin{multline}
\tilde{\psi}_{\mathrm{reg}}(R)=\tilde{\psi}_{0}(R,R)+gF_{1}(R)\tilde{\psi}_{\mathrm{reg}}(R)\\
+g\int_0^{+\infty} d\rho\, F_{2}(R,\rho)\left[\tilde{\psi}_{\mathrm{reg}}(\rho)
-\frac{\rho}{R}\tilde{\psi}_{\mathrm{reg}}(R)\right]
\end{multline}
 that needs to be solved numerically in order to determine $\psi_{\mathrm{reg}}$.
The latter equation can be written in a symbolic, more compact form
as\begin{equation}
\tilde{\psi}_{\mathrm{reg}}=\frac{I}{I-g\hat{O}}\tilde{\psi}_{0}\label{IntegralEq2}\end{equation}
where $\hat{O}$ is a symmetric integral operator, which is real in
the limit $k\rightarrow0$.

If one only aims at calculating strictly zero energy properties (i.e.~$a_{\mathrm{eff}}$,
and not $f_{k}$ or $r_{e}$), the treatment presented in this Appendix
can be drastically simplified setting $\tau=it/\hbar$ 
in Eqs.~(\ref{eq:FreeProp},\ref{eq:HOProp},\ref{eq:GasFourierOfK})
and $E=3\hbar\omega/2$ in Eq.~(\ref{eq:GasFourierOfK}):
the derivation of $\hat{O}$ proceeds in an analogous way, but the
resulting integral equation is much easier to solve since the integrand
in both $F_{1}$ and $F_{2}$ become real damped functions
with no finite time singularities.
For an arbitrary value of the mass ratio $\alpha\equiv m_A/m_B$, 
they are given by:
\begin{multline}
F_1(R)=-\frac{1}{(2\pi)^{3/2}}\int_0^\infty d\tau \left\{ -\frac{1}{\left[\left(1+\frac 1 \alpha\right)\tau \right]^{3/2}}+\right.\\
\left.\frac {\exp(3\tau/2)}{\left[ \mathrm {sh} (\tau) + \frac \tau \alpha \mathrm {ch}(\tau)\right]^{3/2}}
\exp\left(-R^2 \frac {\mathrm {ch}(\tau) + \frac \tau {2\alpha}\mathrm {sh} (\tau)-1}{\mathrm {sh} (\tau) + \frac \tau \alpha \mathrm {ch}(\tau)} \right)\right\},
\label{eq:UF1}
\end{multline}
and
\begin{multline}
F_2(R,\rho)=-\alpha^{3/2} \int_0^\infty d\tau \frac{\exp[-(R^2+\rho^2)x]}{[\tau\mathrm{sh}(\tau)\exp(-\tau)]^{3/2}}\frac{\mathrm{sh}(2R\rho y)}{\left(2\pi\right)^{2}y},
\label{eq:UF2}
\end{multline}
with $x(\tau)=\alpha/2\tau+1/2\mathrm{th}(\tau)$ and  $y(\tau)=\alpha/2\tau+1/2\mathrm{sh}(\tau)$.


\begin{thebibliography}{10}
\bibitem{aPotassium}
If $A$ is a gas of fermionic $^{40}$K atoms ($\lambda_{A}=770$nm) and $B$ a gas of 
$^{87}$Rb, the calculation for $\lambda_L=779$nm yields $V_{0}^{A}=-3.6E_{r}^{A}$. Due to the
smaller detuning and the larger mass ratio $m_A/m_B$, the optical potential would
induce noticeable effects on the potassium matter wave.
\bibitem{does_this_work}
Eqs.~(11,12) of \cite{Dima_loss} are derived under the condition that the kinetic energy of the three particles
{\sl in their center-of-mass frame} is less than the dimer binding energy $\epsilon$:
\begin{equation}
\frac{\hbar^2(k_1^2+k_2^2)}{2m_A}+\frac{\hbar^2 k_3^2}{2 m_B}
-\frac{\hbar^2(\vec{k_1}+\vec{k_2}+\vec{k_3})^2}{2(m_B + 2 m_A)} \lesssim \frac{\hbar^2}{2\mu a_{\rm eff}^2}
\label{eq:ekin}
\end{equation}
where $\mu^{-1}=m_A^{-1}+m_B^{-1}$,
$\vec{k}_1$, $\vec{k}_2$ are the wavevectors of the two $A$ atoms and $\vec{k_3}$ is the wavevector
of the $B$ atom. Whereas low enough values of $k_1, k_2<k_F$ may be chosen by adjusting the density of
$A$ atoms, the value of $k_3$ is fixed by the zero-energy oscillation of the trapped $B$ particle,
$\hbar^2 k_3^2/2 m_B \simeq \hbar\omega$. The $k_1=k_2=0$ limit of Eq.~(\ref{eq:ekin}) therefore
results in a condition that looks difficult to satisfy in general for $a_{\rm eff}\sim d$:
\begin{equation}
\hbar\omega \frac{2m_A}{m_B+2 m_A} \lesssim \frac{\hbar^2}{2\mu a_{\rm eff}^2}.
\label{eq:tough}
\end{equation}
In the limit $m_A\ll m_B$, however, as it is the case for the Li-Rb mixture, the vibrational energy in Eq.~(\ref{eq:tough})
is multiplied by a small number $\sim 2m_A/m_B$ whereas the binding energy
$\epsilon$ increases as $1/m_A$: for $a_{\rm eff}=d=\pi/k_L$ and $V_0^B=50 E_r^B$, we thus find that the condition
is satisfied for $A={}^6$Li, $B={}^{87}$Rb, so that Eqs.~(11,12) of \cite{Dima_loss} may be used.
\bibitem{link_with_T}
The  $T$ matrix of the matter wave is 
related to $M$ by
$T(E+i0^+)=(2\pi\hbar^2/m_A)\sum_{j,l} [M^{-1}]_{jl} |\mathbf{r}_j\rangle \langle \mathbf{r}_l|$.
\bibitem{qed}
This results from the fact that
the Hamiltonian ${\mathcal H}$ is real in position space.
Let $|\psi\rangle\equiv\mathcal{G}(E+i\eta)|\mathbf{r}_{0}\rangle.$
By definition of $\mathcal{G}$, \[
(E+i\eta-\mathcal{H})\psi(\mathbf{r})=\delta\left(\mathbf{r}-\mathbf{r}_{0}\right),\]
where $\cal H$ is a real differential operator.
Taking the limit $\eta\rightarrow0^{+}$ of the imaginary part of
the latter equation leads to $\mathcal{H}\phi=E\phi.$
Alternatively, using the spectral expansion of the Hamiltonian and of the resolvent
with real eigenwavefunctions, one finds 
$\phi(\mathbf{r};\mathbf{r}_0)=-\pi\langle\mathbf{r}|P_E|\mathbf{r}_{0}\rangle $,
where $P_E$ is the projector
on the eigenspace of ${\mathcal H}$ of energy $E$.
This shows that, when one takes all possible values of $\mathbf{r_0}$,
the wavefunctions $\phi$ span the entire eigenspace of energy $E>0$. For $E=0$, the density of states
vanishes in 3D, and so does $\phi$.
\bibitem{bad_case}
This does not hold, however, when $r_+$ or $r_-$ is a real negative number, in which case one recovers the generic
behavior of $\phi(x)$ depicted in Fig.\ref{cap:schema_1d}a. This means that $x_0=0$ is
a `bad' choice, corresponding to the location of a node of the localized state.
\bibitem{details}
We define $\phi_{\textrm{in}}^{2}$ as the average of the 6 values
$\phi^{2}(\mathbf{r}_{\textrm{in}};\mathbf{r}_{\textrm{in}})$, with
$\mathbf{r}_{\textrm{in}}$ chosen between the 6 values $\{(d/2,0,0),(-d/2,0,0),\textrm{ and cyclic permutations}\}$.
Similarly, we define $\phi_{\textrm{out}}^{2}$ as the average of
the 36 values $\phi^{2}(\mathbf{r}_{\textrm{out}};\mathbf{r}_{\textrm{in}})$,
with $\mathbf{r}_{\textrm{out}}$ chosen between the 6 values $\{(20d,0,0),(-20d,0,0),\textrm{ and cyclic permutations}\}$,
and $\mathbf{r}_{\textrm{in}}$ as before.
\bibitem{what_is_xi}
Requiring that the modulus square of the wavefunction decreases by four orders of magnitude
over the radius of the medium (which is a minimal requirement to pretend to be localized)
yields for the parameters of Fig.~\ref{cap:etats3d} a maximal localization length of the order of $4d$.
The numerical observation of larger values of $\xi$ would require a much larger scattering medium.
\bibitem{second_argument} 
We also note that the intuitive fact
that $|m_0(k)|^2$ is minimal in $k=k_0$ leads to a real value for $\beta$.
\bibitem{about_sign} 
Note that $\mbox{Re}\, m_0'$ has to be positive for consistency of the present low-$|m_0|$ theory:
with the reasoning leading to Eq.~(\ref{eq:useful_phi}), we get the approximation for the resolvent
$\langle\mathbf{r}|{\mathcal G}(E+i0^+)|\mathbf{r}_0\rangle \simeq -A\, F(\mathbf{r})F(\mathbf{r}_0)/m_0(k)$.
The linearization procedure in $k$ around $k_0$ leads to $m_0(k)\simeq (\beta m_A/\hbar^2 k_0)\left[E-E_0+i \hbar \Gamma/2\right]$,
where $E=\hbar^2 k^2/2m_A$, $E_0=\hbar^2 k_0^2/2m_A$ and $\Gamma$ is given by Eq.~(\ref{eq:width}).
All the complex poles of the analytic continuation of the resolvent have a negative imaginary part,
so that one should have $\Gamma>0$ and therefore $\beta>0$. This is compatible with the `motion' of the low-imaginary part eigenvalues
of $M$ in the complex plane for increasing $k$, see Fig.\ref{cap:Eigenvalues-of-M}.
\bibitem{perturb}
One can simply use perturbation theory to understand these facts, expanding $M^\infty(k)$
to first order in $k$ and treating the linear terms in $k$ as a perturbation.
\bibitem{why_s}
We restrict to $s$-wave scattering for the $A$ waves, since we expect the $p$-wave
scattering amplitude to vanish generically as $k^2 a_{\rm ho}^3$ at low $k$.
There may be values of $a/a_{\rm ho}$ leading to a $p$-wave resonance, but we assume that
they differ from the locations of the $s$-wave resonances.
\bibitem{better}
One may wonder if one can,
from this formalism suited to the case $m_A\ll m_B$, 
propose an approximate formula giving the value of $a^{-1}$ 
close to the resonance with a negative $a_{\rm res}^{-1}$ and
leading to the desired value of $a_{\rm eff}$. 
We find that the answer is yes if one imposes the constraint
$|a_{\rm ho}/a_{\rm eff}| \ll \alpha^{1/2}$, in which case one has the 
expansion
\begin{equation}
a^{-1}= (1+\alpha) a_{\rm eff}^{-1}-\frac{\sqrt{2}}{a_{\rm ho}} \alpha^{3/2} 
+a_{\rm ho}^{-1} O(\alpha^{5/2}).
\end{equation}
The first term in the right-hand side actually corresponds to the Born approximation.
Because of the imposed constraint on $a_{\rm ho}/a_{\rm eff}$,
this formula may be not very useful in practice.
\bibitem{Wigner_time}
Some reader may fear that the matter wave $A$ spends a very long time 
`inside' a given scatterer because scattering is resonant. However the
Wigner time delay \cite{Wigner}, $\tau(k)=2m_A\delta'(k)/\hbar k$, where $\delta(k)$
is the $s$-wave phase shift, has here a negative value
$\tau\sim -30\hbar/E_r^A$ (for $a_{\rm eff}=d, k a_{\rm eff}=1/3$ and using
the low-$k$ approximation $\delta(k)\simeq -\mbox{atan}\,ka_{\rm eff}$).
\bibitem{discrim}
In particular,
this would allow to discriminate long dwell times due to localization effects from
a parasitic matter wave trapping effect in the scattering medium,
due e.g. to the formation of a weakly bound state between a matter wave atom
and a trapped atom as discussed in \S\ref{subsec:model_ham}.
Note that this particular spurious contribution
to $N_{\rm rem}(\tau)$ could also be eliminated by switching back $a_{\rm eff}$
to a low value just before measuring $N_{\rm rem}(\tau)$, 
so as to transform the weakly bound states into deeply bound states
that can be discriminated from free atoms.
\bibitem{microgravity}Micro-gravity can presently be obtained  e.g.~at ZARM 
(a 120 meters high drop-tower facility in Bremen, Germany, where an experiment with ultracold atoms is planned \cite{Reichel}), 
on airplanes performing parabolic flights
(experiments with atomic clocks on parabolic flights have been reported
in \cite{parabolic98}), or on an orbiting space station (experiments with ultracold atoms on the International Space Station have been
planned by the European Space Agency).%
\bibitem{careful} 
A careful reader may fear that the fraction in the right hand side of Eq.~(\ref{eq:approx_out})
is indeed of large modulus but almost real, so that its imaginary part is actually not
large. As shown in the Appendix \ref{appen:math}, the vector $\vec{v_0}$ is almost real
so that the numerator of the considered fraction is almost real.
\bibitem{a_l_infini}
In the limit $r\rightarrow +\infty$, $F(\mathbf{r})$ involves a factor $e^{ikr}/r$ so it cannot
be real for all $r$. However, using the calculation of the Appendix \ref{appen:magique},
one finds that the angular average of 
$|F(\mathbf{r})|^2$
is of order
$\mbox{Im}\, m_0$, so that the imaginary part of $F(\mathbf{r})$ at large $r$ 
is $O(\mbox{Im}\, m_0)^{1/2}$ and tends to zero with $m_0$.
\bibitem{aide} From Schwartz inequality $|\sum_j x_j \exp(i k\mathbf{n}\cdot\mathbf{r}_j)|^2 \leq
||\vec{x}||^2\times N$. An accurate estimate of the maximal eigenvalue of $M_I$ \cite{Mandonnet} gives a value
$\propto k N/(kL)^2$ in the large $kL$ limit, where $L$ is the diameter of the scattering medium, that is
a value $\propto (kL)^2$ times smaller than the present upper bound.
\bibitem{gaussInt}
$\int_{0}^{\infty}d\rho\,\rho^{2}\exp\left( ix\rho^{2}\right) =\sqrt{\pi/16x^{3}}\exp\left(3i\pi/4\right)$
for $x>0$. The result for $x<0$ is the complex conjugate of the
latter expression.
\end{thebibliography}
\end{document}